\documentclass[11pt, twoside]{article}
\pdfoutput=1
\usepackage{a4wide,amssymb,cite,graphicx}
\usepackage{epsfig}
\usepackage[usenames,dvipsnames]{color}
\usepackage{slashed}
\usepackage{amsmath}

\newcommand{\be}{\begin{equation}}
\newcommand{\ee}{\end{equation}}
\newcommand{\bea}{\begin{eqnarray}}
\newcommand{\eea}{\end{eqnarray}}

\DeclareMathOperator{\sech}{sech}
\begin{document}

\color{black}

\begin{flushright}
CTPU-16-18
 \\
IPMU16-0084
\end{flushright}

\vspace{0.5cm}

\begin{centering}
\vspace{1cm}
{\Large {\bf KK graviton resonance and cascade decays  \\ \vspace{0.2cm} in warped gravity}} \\

\vspace{1.5cm}

{\bf  Barry M. Dillon$^{1,*}$, Chengcheng Han$^{2,\$}$, Hyun Min Lee$^{3,\dagger}$, \\ and Myeonghun Park$^{4,\ddagger}$}

\vspace{.5cm}

%{\it $^1$School of Physics, KIAS, Seoul 130-722, Korea.  }\\
{\it $^1$ Centre for Mathematical Sciences, Plymouth University,
Plymouth, PL4 8AA, UK.  } \\
{\it $^2$ Kavli Institute for the Physics and Mathematics of the Universe (WPI), \\
The University of Tokyo, Japan.} \\
{\it $^3$ Department of Physics, Chung-Ang University, Seoul 06974, Korea.}\\
{\it $^4$ Center for Theoretical Physics and Universe, Institute for Basic Science (IBS), \\
Daejeon 34051, Korea.}\\

\end{centering}
\vspace{2cm}

\begin{abstract}
\noindent
We consider the Kaluza-Klein (KK) graviton and its decays in the presence of the light radion, the modulus of extra dimension, appearing naturally light due to classical scale invariance in warped gravity. Due to the presence of a sizable wavefunction overlap in the extra dimension between the KK graviton and the radion, the KK graviton can decay largely into a pair of radions, each of which decays into a pair of collimated photons or photon-jets.   
Focusing on the setup where the KK graviton has suppressed couplings to the Higgs boson and fermions, we discuss the bounds on the light radion in the parameter space where the photon-jets are relevant. Moreover, we pursue the possibility of distinguishing the scenario from the case with direct photon contributions by the angular distributions of photons in the LHC Run 2. 
Roles of KK graviton and radion as mediators of dark matter interactions,
extensions with brane kinetic terms and the model with a holographic composite Higgs are also addressed.

\end{abstract}

\vspace{1.5cm}

\begin{flushleft}
$^*$Email: barry.dillon@plymouth.ac.uk  \\ 
$^{\$}$Email: chengcheng.han@ipmu.jp  \\
$^\dagger$Email: hminlee@cau.ac.kr \\
$^\ddagger$Email: parc.ctpu@gmail.com  
\end{flushleft}

\thispagestyle{empty}

\normalsize

\newpage

\setcounter{page}{1}

\section{Introduction}

New symmetries in models beyond the Standard Model (BSM) have been searched for via heavy resonances at the LHC, in particular, diboson channels such as diphoton or $ZZ, WW$ could lead to interesting smoking gun signals for BSM. In most BSM models, new gauge bosons, extra Higgs bosons or spin-2 resonances couple to the SM quarks and gauge bosons, decaying directly into the SM particles. However, it is interesting to notice that collimated photons or photon-jets \cite{Knapen,photonjets,Dasgupta,axionmed} can come from the cascade decay of intermediate light states that the heavy resonance decays into. 
Since the efficiency that one or both of collimated photons are converted into an electron positron pair at the LHC can be sizable \cite{Dasgupta}, the photon-jet scenario is soon expected to be distinguishable from the resonances decaying directly into photons.

% our work
In this work, we consider the Kaluza-Klein (KK) graviton \cite{gmdm,tensor,tensor2} and the light radion in the Randall-Sundrum(RS) model \cite{rsmodel}.  However, we also include general couplings on the branes and their effects on the phenomenology of the graviton and radion.  
As  both the KK graviton and the radion are localized towards the IR brane \cite{radion,radion1,radion2}, there is a sizable overlap between their wavefunctions in the extra dimension, so the KK graviton can decay largely into a pair of radions. We assume that the rest of the SM fields, in particular the light fermions, are localized towards the UV brane, in order to avoid strong bounds from dijet and/or dilepton bounds. In principle, the KK graviton could have sizeable couplings to heavy quarks and the Higgs fields. For simplicity, we focus on the case where the KK graviton couples more to the transverse polarizations of gauge bosons than to the rest of the SM fields, but we also mention the case with sizable couplings of the KK graviton to heavy quarks and the SM Higgs fields in the context of the Composite Higgs scenario.  We first discuss the effects of the couplings between KK graviton and radion for the diphoton resonance in a model independent way and then illustrate specific models such as the minimal RS model and its extensions with brane kinetic terms. 

In order for photon-jets to contribute to the diphoton resonance dominantly, we need a large decay branching fraction of the light radion into a photon pair and the radion must decay within the ECAL. In the case that the Yukawa couplings of the radion to light fermions are suppressed due to their localization, a light radion with mass smaller than twice the pion mass can decay mostly into a photon pair. 
In the case with a brane kinetic term for the hypercharge gauge boson, we also show that the coupling of the hypercharge gauge boson to the KK graviton can be reduced such as the KK graviton decays more into a radion pair and the first KK mass of the hypercharge gauge boson is modified to a larger value.

%  organization
The paper is organized as follows.
We begin with a general discussion on the diphoton resonance in the system with a KK graviton and radion and discuss the bounds on a light radion. 
Next we consider specific examples in the RS model and its extensions with brane gauge kinetic terms and discuss the phenomenology of photon-jets. 
Then we take the radion as a mediator of dark matter and discuss the invisible decay of the KK graviton.  We also comment on the case with a holographic composite Higgs model and the LHC bounds.  
Finally, conclusions are drawn.

\section{KK graviton and radion}

We consider the radion, the volume modulus of the extra dimension and/or the KK graviton as good candidates for the diphoton resonance. 
The couplings of the radion and the KK graviton to the SM particles are determined by the wave-function overlaps in the extra dimension. Although the wave functions of the radion and the KK graviton are fixed for a given background geometry, the counterparts of the SM particles are model-dependent.

\begin{figure}
  \begin{center}
    \includegraphics[height=0.55\textwidth]{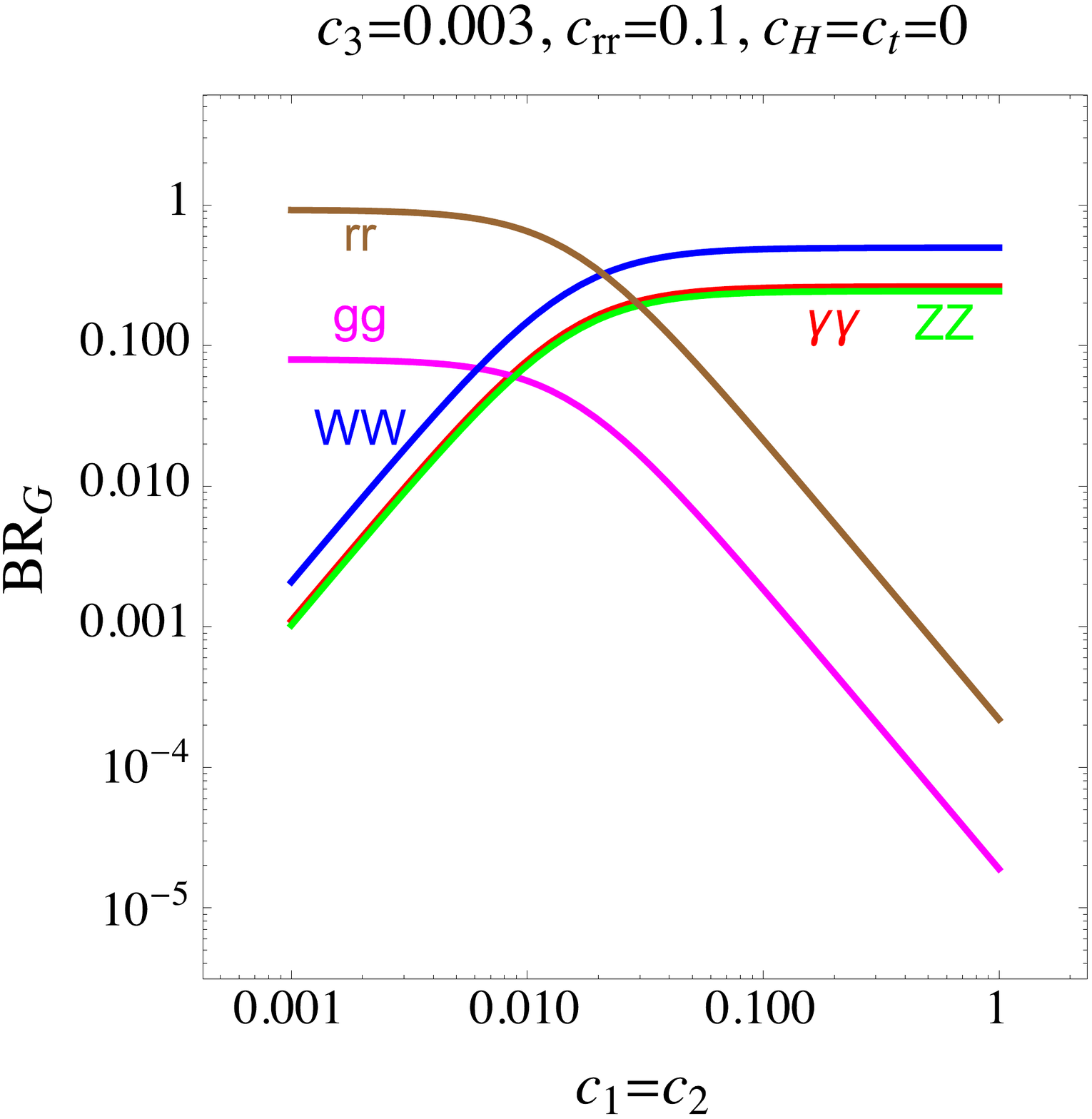}
     \includegraphics[height=0.55\textwidth]{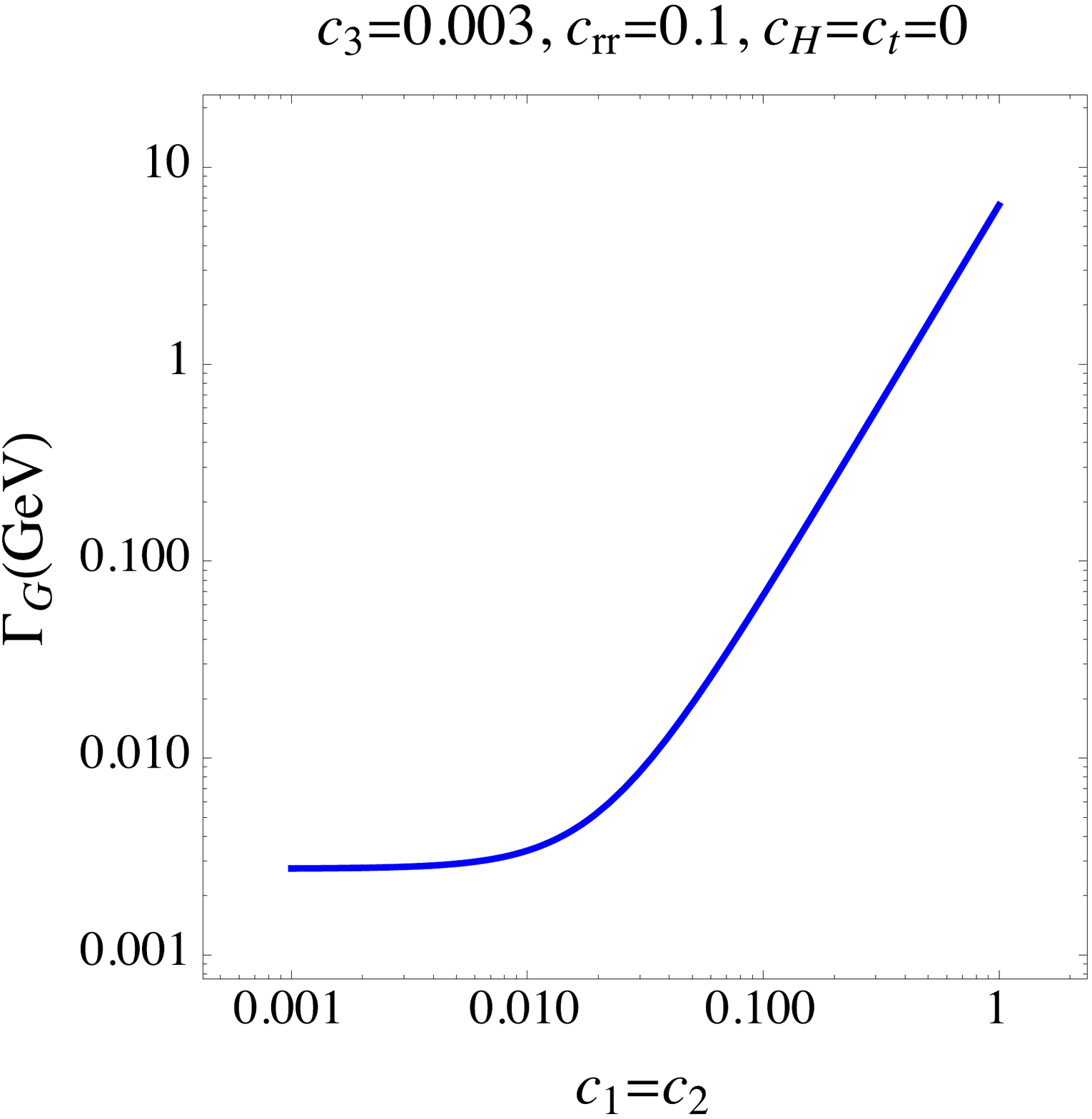}
   \end{center}
  \caption{Decay branching fractions and total decay width of KK graviton as a function of $c_1=c_2$.  We have taken $m_G=1\,{\rm TeV}$, $\Lambda=1\,{\rm TeV}$, $c_3=0.003$ and $c_{rr}=0.1$, and $c_H=c_t=0$. }
  \label{BR-vis}
\end{figure}

We consider the effective interactions of a radion-like spin-0 field, $r$, and a graviton-like spin-2 field, $G_{\mu\nu}$, to SM fields, and their self-interactions, as follows,
\bea
{\cal L}_{\rm eff} &=& \frac{c_1}{\Lambda} G^{\mu\nu} \Big( \frac{1}{4} \eta_{\mu\nu} B_{\lambda\rho} B^{\lambda\rho}+B_{\mu\lambda} B^\lambda\,_{\nu} \Big)+\frac{ c_2}{\Lambda} G^{\mu\nu}  \Big( \frac{1}{4} \eta_{\mu\nu} W_{\lambda\rho} W^{\lambda\rho}+W_{\mu\lambda} W^\lambda\,_{\nu}  \Big)  \nonumber \\
&&+ \frac{c_3}{\Lambda} G^{\mu\nu}  \Big( \frac{1}{4} \eta_{\mu\nu} g_{\lambda\rho} g^{\lambda\rho}+g_{\mu\lambda} g^\lambda\,_{\nu} \Big) \nonumber \\
&& -\frac{ic_f}{2\Lambda}G^{\mu\nu}\left(\bar{f}\gamma_{\mu}\overleftrightarrow{D}_{\nu}f-\eta_{\mu\nu}\bar{f}\gamma_{\rho}\overleftrightarrow{D}^{\rho}f\right) \nonumber \\
&&+\frac{c_H}{\Lambda}G^{\mu\nu}\left(2(D_{\mu}H)^{\dagger}D_{\nu}H-\eta_{\mu\nu}\left((D_{\rho}H)^{\dagger}D^{\rho}H-V(H)\right)\right) \nonumber \\
&&-\frac{r}{\Lambda}\,  \Big(d_1B_{\lambda\rho} B^{\lambda\rho}+ d_2W_{\lambda\rho} W^{\lambda\rho}+ d_3 g_{\lambda\rho} g^{\lambda\rho}  \Big) \nonumber \\
&&+\frac{d_f}{\Lambda}\, m_f r {\bar f}f+\frac{4d_H}{\Lambda}\, m^2_H r |H|^2-\frac{d_V}{\Lambda}\, r (m^2_Z Z_\mu Z^\mu+ 2 m^2_W W_\mu W^\mu) \nonumber \\  
&&-\frac{c_{rr}}{\Lambda} \bigg[ {G}^{\mu\nu}\partial_\mu{r}\partial_\nu {r} + \frac{3}{2}{G}(\partial { r})^2 +{r}^2(\partial_\mu\partial_\nu {G}^{\mu\nu}-\Box { G}) \bigg]   \label{lag}
\eea
where $B_{\mu\nu}, W_{\mu\nu}, g_{\mu\nu}$ are the strength tensors for $U(1)_Y, SU(2)_L, SU(3)_C$ gauge fields, respectively, $f$ is the SM fermion, $H$ is the Higgs doublet.
Here, we note that $c_{i}(i=1,2,3)$, $c_f$, and $c_H$ are constant dimensionless couplings for the KK graviton while $d_i(i=1,2,3)$, $d_f$,  $d_H$, $d_V$, and $c_{rr}$ are those for the radion.  The couplings between KK graviton and radion are derived from the 5D Einstein-Hilbert action and the detailed derivation of those gravity self-interactions in the 5D warped geometry can be found in the appendix.

\begin{figure}
  \begin{center}
          \includegraphics[height=0.55\textwidth]{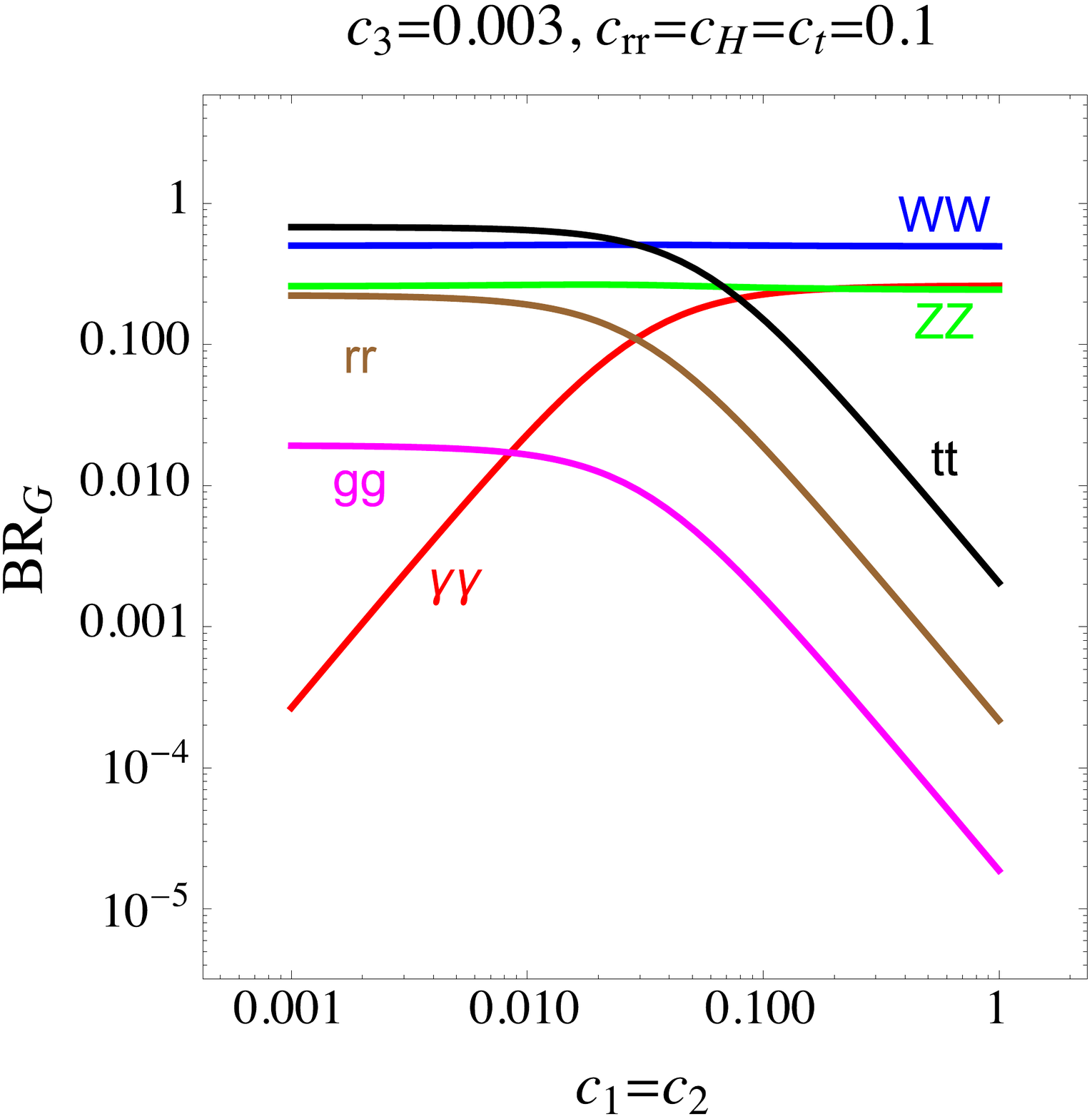}
     \includegraphics[height=0.55\textwidth]{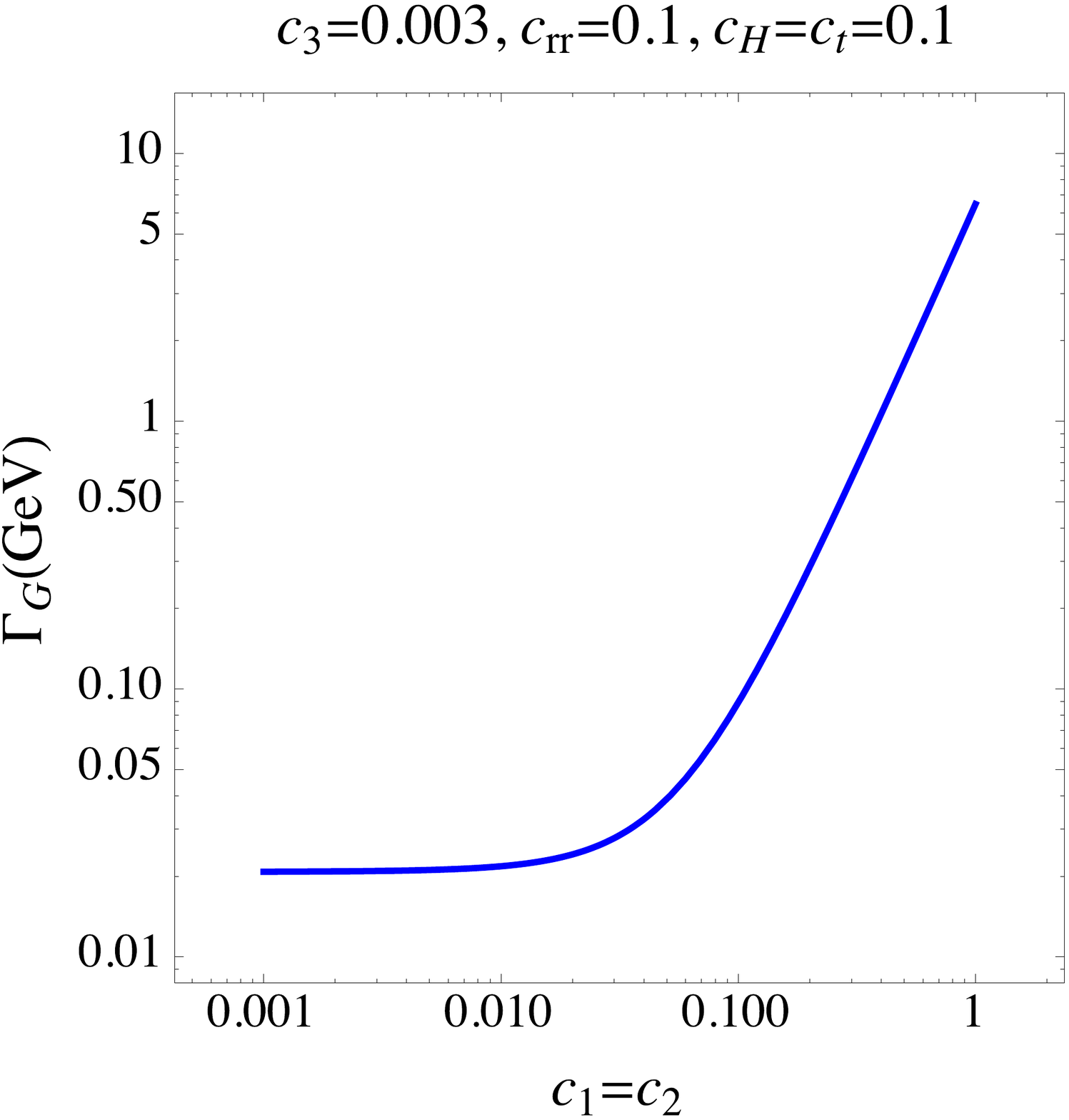}
   \end{center}
  \caption{The same as Fig.~\ref{BR-vis} but with $c_H=c_t=0.1$. }
  \label{BR-vis2}
\end{figure}

From the effective interactions given in eq.~(\ref{lag}), the partial decay rates of the KK graviton \cite{gmdm} are given by
\bea
    \Gamma_G(gg)&=& \frac{c_{gg}^2m_G^3}{10\pi \Lambda^2}, \quad\quad
       \Gamma_G(\gamma\gamma)=\frac{c_{\gamma\gamma}^2m_X^3}{80\pi \Lambda^2},  \nonumber \\
      \Gamma_G(ZZ)&=&\frac{m_G^3}{80\pi \Lambda^2}\sqrt{1-4r_Z}\bigg(c_{ZZ}^2+\frac{c_H^2}{12}+\frac{r_Z}{3}\left(3c_H^2-20c_Hc_{ZZ}-9c_{ZZ}^2\right) \nonumber \\
    &+&\frac{2r_Z^2}{3}\left(7c_H^2+10c_Hc_{ZZ}+9c_{ZZ}^2\right) \bigg),   \nonumber  \\\Gamma_G(WW)&=&  \frac{m_G^3}{40\pi \Lambda^2}\sqrt{1-4r_W}\bigg(c_{WW}^2+\frac{c_H^2}{12}+\frac{r_W}{3}\left(3c_H^2-20c_Hc_{WW}-9c_{WW}^2\right) \nonumber \\
    &+&\frac{2r_W^2}{3}\left(7c_H^2+10c_Hc_{WW}+9c_{WW}^2\right)\bigg),   \nonumber \\
    \Gamma_G(Z\gamma)&=& \frac{c_{Z\gamma}^2m_G^3}{40\pi \Lambda^2}(1-r_Z)^3\left(1+\frac{r_Z}{2}+\frac{r_Z^2}{6}\right),   \nonumber 
\eea
\bea
    \Gamma_G(f\bar{f})&=&\frac{N_c c_f^2 m_G^3}{320\pi \Lambda^2} (1-4r_f)^{3/2}(1+8r_f/3),  \nonumber \\
      \Gamma_G(hh)&=&\frac{c_H^2m_G^3}{960\pi \Lambda^2}(1-4r_h)^{5/2},  \nonumber \\
  \Gamma_G(rr) &=& \frac{c^2_{rr} m^3_G}{960\pi \Lambda^2}\left(1-4r_r \right)^{5/2}
\eea
where $c_{\gamma\gamma}=s_{\theta}^2c_2+c_{\theta}^2c_1$, $c_{ZZ}=c_{\theta}^2c_2+s_{\theta}^2c_1$, $c_{Z\gamma}=s_{\theta}c_{\theta}(c_2-c_1)$, $c_{gg}=c_3$, $c_{WW}=2c_2$, $r_i=(m_i/m_G)^2$, and $m_G$ is the lightest KK graviton mass.

\begin{figure}
  \begin{center}
    \includegraphics[height=0.40\textwidth]{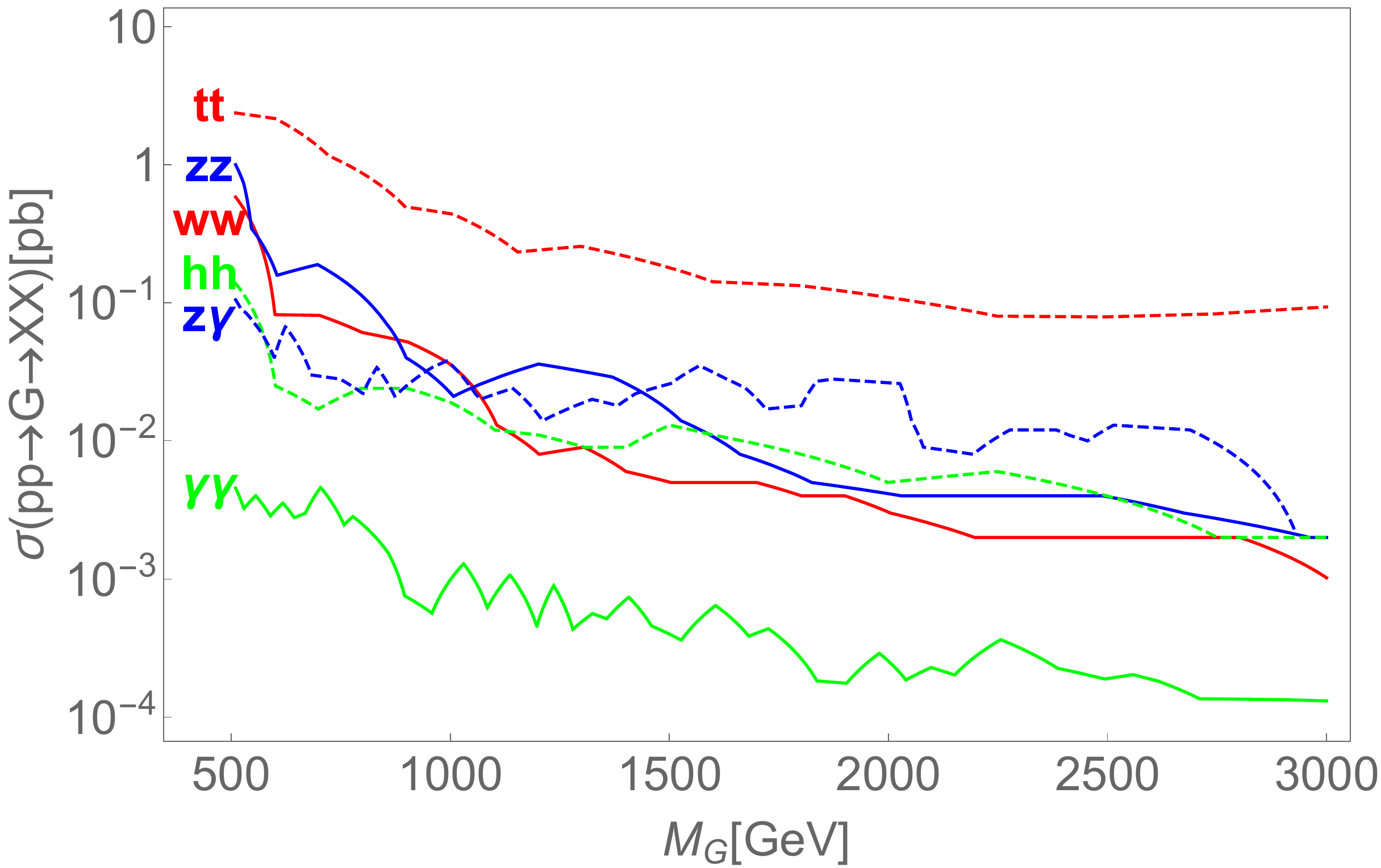}
      \end{center}
  \caption{Limits on the production cross sections for decay channels of the KK graviton at LHC 13TeV. }
  \label{LHCmasses}
\end{figure}

In Fig.~\ref{BR-vis}, we depict the decay branching fractions and total decay rate of the KK graviton as a function of $c_1=c_2$ by assuming that the Higgs fields and the SM fermions have suppressed couplings to the KK graviton, namely, $c_H=c_f=0$ in the upper panel (which is the main focus of our paper) and $c_H=c_t=0.1$ in the lower panel (for the model with a holographic composite Higgs). 
We took $c_3=0.01$ and $c_{rr}=0.1$ and assumed that there is no invisible decay of the KK graviton. Then, the KK graviton decays into $WW, ZZ, \gamma\gamma, gg$ or $rr$. In the case with $c_H=c_t=0.1$, there are additional decay modes such as $hh, t{\bar t}$. 
The $rr$ mode is relevant for the cascade decay of the KK graviton in a later discussion. 
In a later section, we will also comment on the effects of the invisible decay into a pair of dark matter.

The current upper bounds on the partial cross sections of the KK graviton with $m_G=1\,{\rm TeV}$  at LHC 13TeV, apart from $t{\bar t}$ at LHC 8TeV, are obtained from Refs.~\cite{Aaboud:2017yyg,ATLAS:2017xvp,ATLAS:2016npe,Aaboud:2016trl,ATLAS:2016ixk,Aad:2015fna}, as shown in Table~\ref{LHCbounds}.
Moreover, we also include the cross section limits on various decay channels of the KK graviton in Fig~\ref{LHCmasses}. 
\begin{table}[htp]
\caption{LHC bounds for KK graviton with $m_G=1\,{\rm TeV}$ on the differential channels.}
\begin{center}
\begin{tabular}{|c|c|c|c|c|c|c|c|c|}
\hline
Channels & $\gamma\gamma$ \cite{Aaboud:2017yyg}& $WW$ \cite{ATLAS:2017xvp} & $ZZ$ \cite{ATLAS:2016npe} & $Z\gamma$ \cite{Aaboud:2016trl} & $hh$ \cite{ATLAS:2016ixk}& $t\bar{t}$ \cite{Aad:2015fna} \\
 \hline  Limit(fb) & 1.0 & 35 & 20 & 36 & 19 &  433 \\
 \hline
\end{tabular}
\end{center}
 \label{LHCbounds}
\end{table}%

On the other hand, the decay modes of a light radion \footnote{The possibility of having a heavy radion was also recently considered \cite{diphotonradion}.}  are
\bea
\Gamma_r(\gamma\gamma) &=&\frac{d^2_{\gamma\gamma} m^3_r}{4\pi \Lambda^2}, \\
\Gamma_r(gg) &=&  \frac{2d^2_{gg} m^3_r}{\pi \Lambda^2}, \\
\Gamma_r({\bar f}f)&=& \frac{d^2_f m^3_f}{8\pi \Lambda^2} \Big(1-\frac{4m^2_f}{m^2_r} \Big)^{1/2}
\eea
where
$d_{\gamma\gamma}= d_1\cos^2\theta_W + d_2 \sin^2\theta_W$ and
$d_{gg}= d_3$.

When the radion is lighter than $2m_\pi<m_r\lesssim 1.5\,{\rm GeV}$, we need to include the radion decays into a pair of mesons in chiral perturbation theory, instead of the decay into a gluon pair. For instance, the decay rate of the radion into a pair of pions  ($\pi^a\pi^a$ with $a=1,2,3$)  is
\be
\Gamma_r(\pi^a\pi^a)=\frac{m^3_r}{32\pi\Lambda^2} \left[\frac{d_f m^2_\pi}{m^2_r} +\frac{8\pi d_{gg}}{b\,\alpha_S} \Big(1+\frac{m^2_\pi}{m^2_r}\Big)  \right]^2
\,\sqrt{1-\frac{4m^2_\pi}{m^2_r}}  \label{pions}
\ee 
with the coefficient of the QCD beta function being $b=\frac{29}{3}(9)$ for $u,d (u,d,s)$ light quarks in chiral perturbation theory.
Here, the radion effective gluon coupling, $d_{gg}(M_Z)$, and the strong gauge coupling, $\alpha_S(M_Z)=0.118$, can be chosen at $Z$ pole, near the matching scale to the extra dimension. 
We note that the loop contributions of heavy quarks to the radion couplings are thought as being absorbed into the bulk radion coupling, that is taken to be dominant in the following discussion.

\begin{figure}
  \begin{center}
    \includegraphics[height=0.40\textwidth]{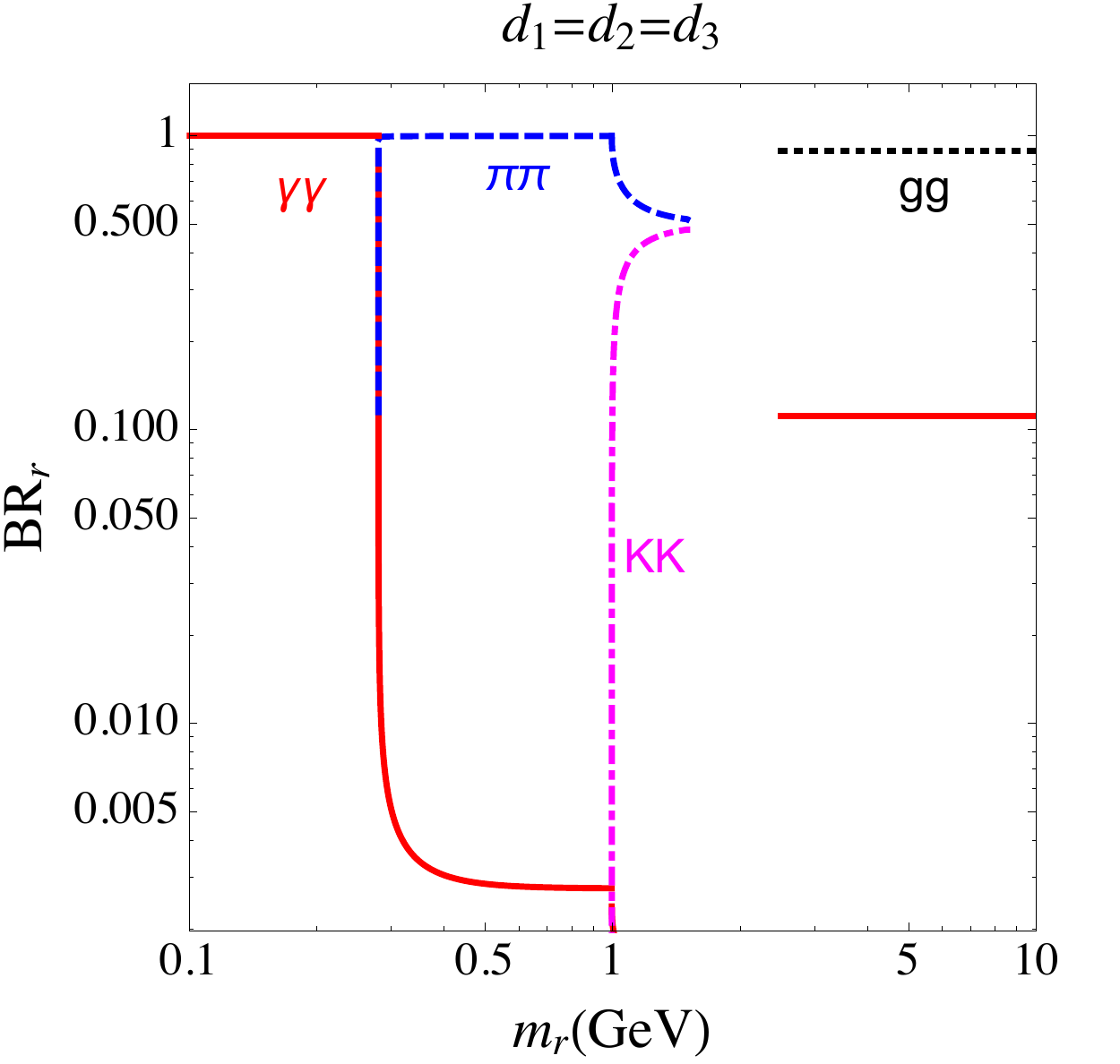}
     \includegraphics[height=0.40\textwidth]{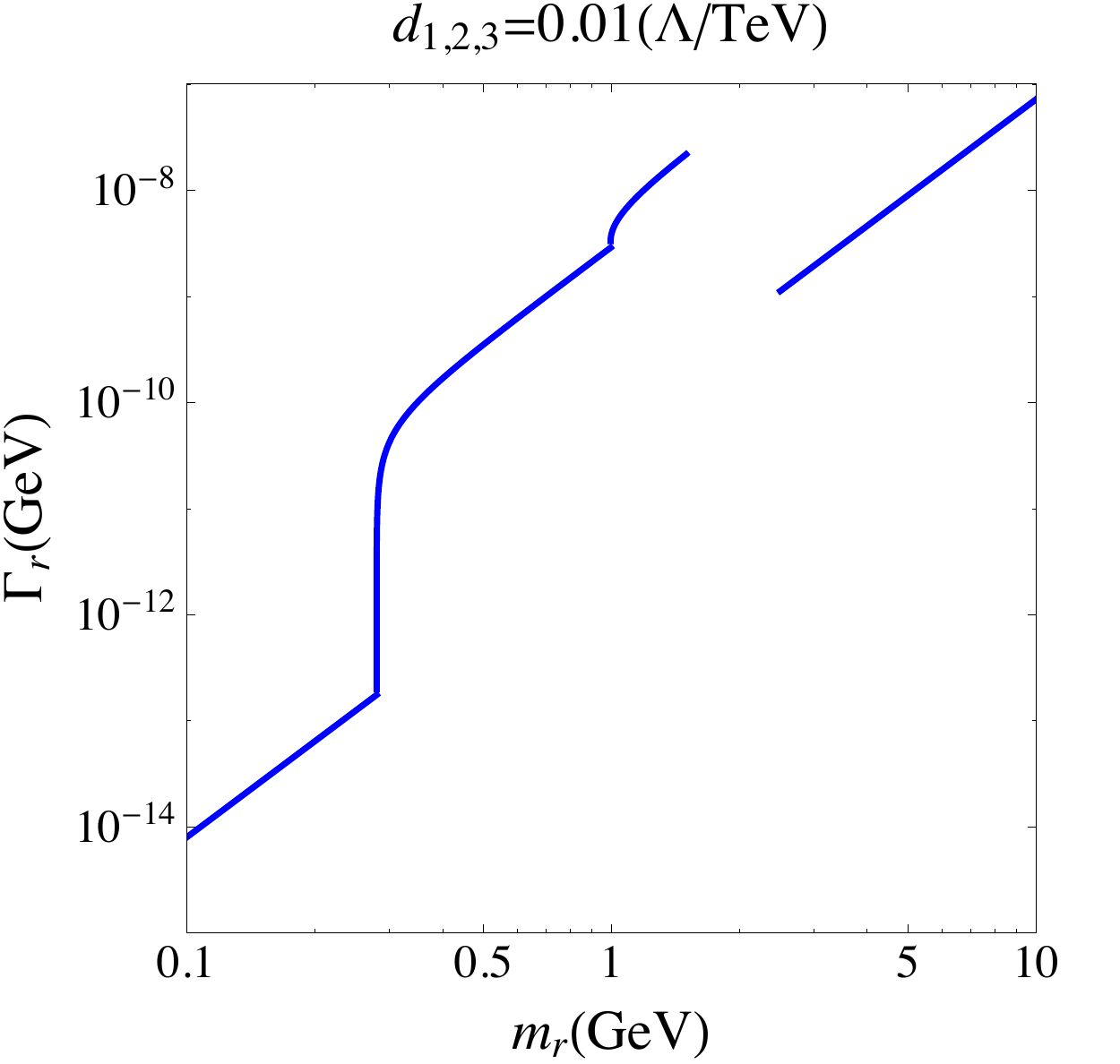}
   \end{center}
  \caption{Decay branching fractions and total decay width of radion as a function of radion mass.  We have taken $d_1=d_2=d_3=0.01(\Lambda/{\rm TeV})$. We set the fermion couplings to zero, i.e. $d_f=0$. }
  \label{BR-radion}
\end{figure}

In Fig.~\ref{BR-radion}, we show the decay branching fractions and total decay rate of the radion by assuming that the Higgs fields and the SM fermions have suppressed couplings to the KK graviton, namely, $d_H=d_V=d_f=0$.  Here, we have chosen $d_1=d_2=d_3=0.01(\Lambda/{\rm TeV})$ for $\Lambda=1\,{\rm TeV}$. We note that the branching fractions or the total decay rate of the radion are not shown for $1.5\,{\rm GeV}<m_r<2.5\,{\rm GeV}$, for which either descriptions in terms of mesons or quarks/gluons are not quite correct.

In the following discussion, we consider the case that the spin-2 resonance couples more strongly to transverse polarizations of gauge bosons than to the rest of the SM fields and focus on photon and gluon couplings to the spin-2 resonance after electroweak symmetry breaking.

\section{Radion and KK graviton from warped geometry}

In this section we consider a concrete model for the effective interactions of the radion and a KK graviton in the context of warped gravity, in particular, the RS model without or with brane gauge kinetic terms. 
We discuss some benchmark models for the decays of KK graviton and radion where the SM fields apart from gauge fields are localized on the UV brane.
Then, in Section 3.3, we show how brane gauge kinetic terms affect the radion couplings in these benchmark models. 
Later in Section 6, we will also consider the case where the Higgs and  third generation quarks are localized on the IR brane.

\subsection{Wave functions of radion and KK graviton}

We identify the Kaluza-Klein graviton and the radion in the RS framework  as CP-even spin-2 and spin-0 fields, respectively, discussed in the previous section. The corresponding 5D gravitational action is given by
\be \label{5Daction}
S_5=\int d^4x\int_0^L dy~\sqrt{|g|} M_{*}^3\left(-\mathcal{R}_5+12k^2 \right) +\left( \sqrt{|g|}M_{*}^3\left(k_B\pm\frac{1}{2k}r_{0,L}\mathcal{R}_4\right)\right)\Big\vert^L_0
\ee
where $M_{*}$ is a mass term which defines the 5D fundamental scale, $k$ is the bulk curvature constant, and $k_B$ determines the brane tensions which allow an AdS solution to the Einstein equations for the bulk.  The 4D Einstein terms on the branes generate brane kinetic terms (BKTs) for the graviton and radion, controlled by the $r_{0,L}$ parameters.  The resulting metric is that of the Randall-Sundrum model,
\be  \label{RSmetric}
ds^2 = e^{-2ky} \eta_{\mu\nu} dx^\mu dx^\nu - dy^2.
\ee
At $y=0$ and $y=L$ we have the UV and IR 3-branes, and the mass scale at which the Kaluza-Klein (KK) modes appear at is approximately $ke^{-kL}=M_{KK}$.

Our primary interest consists in models in which all of the SM fields are localised on the UV brane, with the exception of the SM gauge fields, which are in the bulk.   In the RS models it is expected that the lightest spin-2 KK mode is heavier than the lightest spin-1 mode, where the lower bounds on the latter are $\sim 3$ TeV.  However, this mass constraint can be avoided if appropriate gravity BKTs are introduced as in our action, eq.~(\ref{5Daction}).  The large values of $r_L$ usually would lead to a negative effective kinetic term for the radion, which could be avoided due to additional IR dynamics \cite{Dillon:2016bsb}.
Effects of this gravity BKT have also been studied recently in the context of holographic phase transitions \cite{holoPhase}.

Perturbing around the RS background (\ref{RSmetric}), we find the graviton fluctuations, associated with the 4D components of the metric, and the radion, identified with the fluctuations along the $(55)$ component of the metric.  With the above considerations in mind, the 5D wavefunctions for the radion and graviton are found to be
\bea
\hat{r}(x,y)=f_r(y) r(x), \nonumber \\
\hat{G}_{\mu\nu}(x,y)=f_G^{(n)}(y)G_{\mu\nu}^{(n)}(x)  \label{profiles}
\eea  
with
\bea
f_r(y)&\equiv&\frac{1}{\sqrt{6}\Lambda_r}e^{2k(y-L)}, \\
 f_G^{(n)}(y)&\equiv&\frac{e^{2ky}}{N_n}\left( J_2\left(\frac{m_ne^{ky}}{k}\right) - \beta_n(m_n)Y_2\left(\frac{m_ne^{ky}}{k}\right)  \right).
\eea
Here, $m_n$ are the KK masses of the 4D modes, $\beta_n$ is an integration constant, and $N_n$ is a constant fixed by requiring that the kinetic term is normalised,
\be
\frac{1}{k}\int_0^Ldy~e^{-2ky}f^{(m)}_G f^{(n)}_G\left(k+r_L\delta(y-L)\right)=\frac{4\delta_{mn}}{M_{*}^3}.
\ee
The integration constant and the KK masses are fixed by the boundary conditions for the graviton on the UV and IR branes.  The boundary conditions for the graviton fields can be written as
\begin{align}
f_G^{(n)}(0)=&~0 \nonumber \\
f_G^{(n)\prime}(L)=&\frac{r_Lm_n^2}{ke^{-2kL}}f_G^{(n)}(L).
\end{align}
We then find the integration constant to be
\be
\beta_n(m_n)=\frac{J_2(m_n/k)}{Y_2(m_n/k)}\ll1.
\ee
In the presence of the IR BKT there exists an additional light mode in the spectrum with mass,
\be
m_1\simeq\frac{2M_{KK}}{\sqrt{r_L}},
\ee
hence using $r_L$ we can fix the first graviton mass.
As a consequence, we can make the light KK graviton lighter than KK modes of bulk gauge bosons with masses $\gtrsim M_{KK}$, although it is possible at the expense of the 4D Einstein terms.

\subsection{Couplings in the RS model}

In this section we discuss the couplings of radion and KK graviton in the minimal RS model. 

First, the effective couplings of the KK graviton to the SM fields are given by the integration over  the 5D profiles of gauge fields and KK graviton,
\begin{align} \label{gravitonGaugeHiggs}
c_{1,2,3}=&\frac{\Lambda}{2L}\int_0^Lf_G^{(1)}dy \nonumber \\
c_{f,H}=&\frac{\Lambda}{2}\int_0^Lf_{f,H}^2f_G^{(1)}dy,
\end{align}
where $f_{f,H}$ denotes the 5D profile of the fermion with $f=(q,tR,bR)$ or Higgs field and $\Lambda$ is the IR cutoff scale, which is given by $\Lambda=(M_{*}/k)^{3/2}M_{KK}=M_P\, e^{-kL}$ and the same as $\Lambda_r$  in the minimal RS model.  We note that the effective couplings of the KK graviton to the bulk SM gauge fields are suppressed by the volume of the extra dimension, such as $(kL)^{-1}$
The 5D profiles of bulk SM fields can be written in a compact form, 
\be
f_a=\sqrt{\frac{2ak}{1-e^{-2akL}}}e^{-2aky}
\ee
such that they are normalised according to $\int_0^Ldyf_a^2=1$ and $a=(a_H,a_{q},a_{tR},a_{bR})$ represents the localisation of the 5D fields.  

The couplings of the radion to the transverse gauge fields are,
\be \label{radionGauge}
d_i=\frac{\Lambda}{\Lambda_r\sqrt{6}}\left(\frac{1}{4kL}+\frac{\alpha_i}{8\pi}(b_i+F_i)\right), \quad i=1,2,3
\ee
where $b_i$ are the beta function coefficients of the light fields localized on the IR brane, and $F_i$ are loop functions coming from the couplings to heavy fields such as $W$ and top quark \cite{radion2}.  If we have the Higgs and fermions on the UV brane, thus only having the gauge fields, KK gravitons, and radion in the bulk, then the couplings of the KK graviton to the Higgs and fermions are negligible as compared to the couplings to the transverse gauge fields.  In addition to this, with a radion of mass $\sim 0.2$ GeV and the heavy KK graviton, the only relevant decay channels are $G\rightarrow VV$, $G\rightarrow rr$, and $r\rightarrow VV$.  If we were to have the Higgs and some fermions localised near the IR then the situation would change drastically.  The KK graviton would then have large branching fractions to the Higgs, the IR localised fermions, and the longitudinal components of $W^{\pm}$ and $Z$.

There is also a triple coupling between KK graviton and two radions\footnote{There is a triple coupling between two KK gravitons and one radion too, as shown in the appendix, but it is not relevant for our discussion on the diphoton resonance. } stemming from the graviton self-interactions of the 5D Einstein term.
As shown in the appendix, this is also similarly given by the integration over the 5D profiles of KK graviton and radion as 
\be
c_{rr}=\frac{\Lambda}{3\Lambda_r^2}\int_0^Ldy~e^{2k(y-2L)}f^{(1)}_G\simeq\frac{\Lambda}{3N_1\Lambda_r^2}\int_0^Ldy~e^{4k(y-L)}J_2\left(\frac{m_1e^{ky}}{k}\right),
\ee
where the last expression naively assumes that there are no gravity BKTs.  
We note that the above radion coupling is independent of the volume of the extra dimension, that is, $(kL)^{-1}$, unlike the effective couplings of the KK graviton to the SM gauge fields in bulk. 
Assuming that $m_r\ll m_G^{(1)}$, the above couplings should be the most relevant graviton self-interaction.  
The KK graviton couplings to two radions is given numerically by $c_{rr}\sim 0.1$, which is sizable due to the simultaneous localization of both KK graviton and radion toward the IR brane.  Therefore, such a sizable coupling could lead to the observable diphoton resonance in terms of the cascade decay of the KK graviton into a pair of radions, with each radion decaying into a pair of photons.

One could also introduce the non-minimal $\xi \mathcal{R}H^{\dagger}H$ coupling in our discussion, which induces a mixing between the radion and the Higgs as well as additional BKT terms for the radion and graviton \cite{radion2,diphotonradion}.  If the Higgs is localised on the UV brane, this coupling will be very suppressed due to the small overlap between the Higgs and the radion.  With an IR localised Higgs the effect could be large, however it is interesting to note that in the scenario where the Higgs arises as a Goldstone boson, the non-minimal coupling cannot be induced at tree-level, and is expected to be small \cite{Cox:2013rva}.  In our work we will neglect this Higgs-radion mixing.

Henceforth, we focus on the case that the SM fields apart from gauge fields are localized on the UV brane.
In this case,  without graviton BKTs, the gauge boson couplings to the 1st KK graviton are $c_i\sim (kL)^{-1}= [\ln(\Lambda_{\rm UV}/\Lambda)]^{-1}\sim0.03$ with an IR scale of $\Lambda\sim 1\,{\rm TeV}$ and a UV scale of $\Lambda_{UV}=M_P$.  And at the same time, ignoring trace anomalies and loop corrections, we obtain the gauge boson couplings of the radion as $d_i\sim (kL)^{-1}/(4\sqrt{6})= [4\sqrt{6}\ln(\Lambda_{\rm UV}/\Lambda)]^{-1}\sim 0.003$.  Therefore, in the minimal setup, the radion couplings to gauge bosons are much suppressed as compared to those of the KK graviton.  The minimal couplings of radion and KK graviton in RS model (without gravity BKTs, i.e. $r_L=0$) are summarized in Table~\ref{tab2}.  In the next section, we discuss the effects of BKTs of gauge fields on the radion couplings to gauge bosons. 

\begin{table}[th] 
\begin{tabular}{|c|c|c|c|c|c|c|c|}
\hline      & VV & rr  & GG \\
\hline  radion   & $d_i=0.003\,(35/(kL))$   &  $-$ & $c_{GG}=0.2$    \\
\hline KK graviton  & $c_i=0.03\, (35/(kL))$  & $c_{rr}=0.1$ & $-$ \\
\hline
\end{tabular}
\centering\caption{Minimal couplings of radion and KK graviton in RS model as a function of the volume factor, $kL$.}
\label{tab2}
\end{table}
%%%%%%%%%%%%%%%%%%%%%%%%%%%%%%%%%%%%%%%%%%%%%%%%%%%%%%%%%%%%%
From the couplings from Table~\ref{tab2},  the KK graviton decays dominantly into a pair of gluons while the decay rate of the KK graviton into a pair of radions is comparable to the one into a photon pair.  
On the other hand, for the radion with $m_r<2m_{\pi}\approx 280\,{\rm MeV}$, the radion decays fully into a pair of photons with the corresponding decay rate,
\begin{eqnarray} \label{BFradion}
\Gamma_{r}(\gamma\gamma)=\frac{d_{\gamma\gamma}^2 m_r^3}{4\pi \Lambda^2}< 8.95\times10^{-14}~\textrm{GeV} \Big(\frac{1\,{\rm TeV}}{\Lambda}\Big)^2\Big(\frac{35}{kL}\Big)^2.
\end{eqnarray}
Then, the would-be photon-jet contribution to the diphoton cross section, $R=\Gamma_G(rr)/\Gamma_G(\gamma\gamma)=0.93(kL/35)^2$. However, in order for the radion to decay inside the ECAL as will be discussed in the next section, we would need $kL\lesssim 11$, which implies that the UV scale must be $\Lambda_{\rm UV}\lesssim 6\times 10^7\,{\rm GeV}$ for $\Lambda=1\,{\rm TeV}$. But, for $kL\lesssim 11$, the KK graviton couplings would increase to $c_i\gtrsim 0.1$, so the $R$-ratio gets a smaller value, $R\lesssim 0.1$. 
Therefore, the minimal RS model with light radion could not give rise to significant photon-jet signals. In this case, the single photon contribution from the direct decay of the KK graviton would be dominant, as shown in Ref.~\cite{tensor}. 
If $kL\gtrsim 11$, the radion would decay outside the ECAL such that it could be regarded as missing momentum at the LHC.

\subsection{Brane gauge kinetic terms}

In this section, we show that the couplings of the KK graviton to the gauge fields can be made suppressed in the presence of brane kinetic terms (BKTs) for gauge fields, while the radion couplings to them remain unaffected due to the classical scale invariance of those BKTs.  
In this case, significant photon-jet contributions can be achieved. 
We also discuss the effects of gauge BKTs on the non-universal couplings of the KK graviton as well as the lightest KK masses of gauge fields.

Since both the KK graviton and radion couplings to the gauge fields are proportional to the inverse volume of the warped extra dimension, $(kL)^{-1}$, we can take a smaller $kL$ in the little RS model to enlarge the radion couplings while maintaining or reducing the KK graviton couplings at the expense of gauge BKTs. In this case, the radion can decay within ECAL without affecting the ratio $R$ for the photon-jet contribution found in the RS model in the previous section.  Therefore, the photon-jet scenario is valid due to the non-minimal couplings.

Consider the following action for a bulk gauge field in the warped geometry with a BKT,
\be
S=-\frac{1}{4}\int d^4x\int_0^L\sum_{a=1}^3\Big(F^{a, MN} F^a_{MN}+\sigma_a F^{a, \mu\nu} F^a_{\mu\nu}\delta(y-L)\Big),
\ee
where we do not introduce a BKT for the $(\mu5)$ components of the field strength tensor as they are $Z_2$-odd, thus ignore them here.  The BKTs, $\sigma_a$, are dimensionful, and modify the normalisations of the zero mode wave functions such that,
\be
A_{a,\mu}(x,y)=\frac{A_{a,\mu}^0(x)}{\sqrt{L+\sigma_a}} + \ldots.
\ee
By inspecting the interactions of the gauge fields we see that the relation between the 5D and effective gauge couplings in turn gets modified,
\be
(g^a_5)^2=(g^a_4)^2(L+\sigma_a).
\ee
Since we fix $g^a_4$, this parameter modifies the couplings of localised fields which couple via $g^a_5$, i.e. KK modes.  The couplings of the KK graviton and the radion get modified by this BKT in two ways; one being through the normalisation of the gauge field, and the other being due to the direct coupling of the KK graviton to the BKT.  In these warped models we find that the dimensionless couplings for the KK graviton in the effective theory are given by,
\be \label{gravitonGaugeHiggsBKT}
\frac{c_a}{\Lambda}=\frac{1}{2}\frac{1}{\sqrt{L+\sigma_a}}\int_0^L(1+\sigma_a\delta(y-L))f_G^{(1)}
\ee
and the dimensionless couplings of the radion are given by,
\be  \label{radionGaugeBKT}
\frac{d_a}{\Lambda}=\frac{(k/M_*)^{3/2}}{\sqrt{6}M_{KK}}\left(\frac{1}{4k(L+\sigma_a)}+\frac{\alpha_a}{8\pi}(b_a+F_a)\right).
\ee

\begin{figure}
  \begin{center}
    \includegraphics[height=0.40\textwidth]{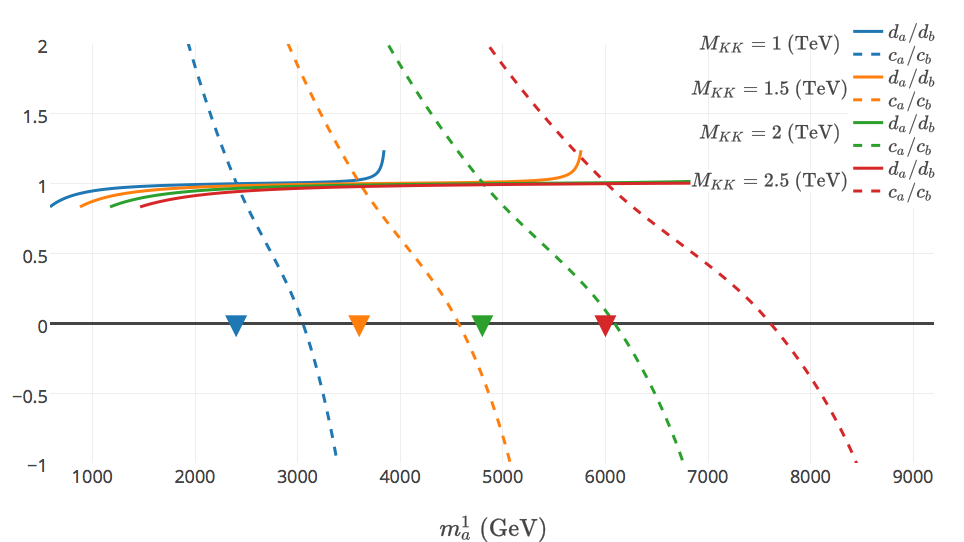}
   \end{center}
  \caption{Here we plot the effects of gauge BKTs on the couplings of the gauge fields to KK graviton and radion, and the triangles denote the masses of the first $A_{\mu}^a$ mode when the BKT is zero.  The BKT for $A_{\mu}^a$ is varied over $-0.2L<\sigma_a<0.2L$ and the BKT for $A_{\mu}^b$ is kept zero.  Note that to keep the results general, we have ignored the effects of the anomalous terms.\label{BKTgaugeCouplingRatios}}
  \label{kkmass}
\end{figure}

As well as modifying the couplings of the gauge fields, the BKTs also modify the mass spectrum of the spin-1 modes, in fact the zero mode remains in the spectrum and the lightest massive mode is given by,
\be
m^{1}_{A^a}\simeq 2M_{KK}\sqrt{\frac{kL+\sigma_a k}{kL-1+2\sigma_a k^2L-\sigma_a k}}.
\ee
If the BKTs become too large we would inevitably have light spin-1 fields in the spectrum which may be in tension with experimental bounds.

In Fig.~\ref{BKTgaugeCouplingRatios}, to quantify the effects of these BKTs, we plot ratios of couplings ($c_a/c_b$ and $d_a/d_b$) to different gauge fields ($A_{\mu}^a$ and $A_{\mu}^b$) where we add a BKT ($\sigma_a$) only for the $A_{\mu}^a$ gauge field, varying over $-0.2L< \sigma_a<0.2L$.  In this case, we note that there is no instability due to negative mass squared or negative norm states. 
To make this analysis general, we neglect the anomalous contributions to the radion couplings.
As a result, we find that one can completely suppress the graviton coupling to the transverse gauge fields ($c_a/c_b$), while the radion coupling to transverse gauge fields ($d_a/d_b$) remains mostly unaffected.  
Even for a small negative BKT for the bulk gauge fields, we can suppress the graviton coupling to photons while at the same time increasing the lightest KK photon mass, therefore the dominant di-photon production can be due to the photon-jets coming from radion decays.  How large the photon-jet contribution is only depends on how much we are willing to tune the BKT.  Note that we have checked that the negative BKTs do not lead to large changes in the effective kinetic terms for the gauge fields, the effect on the KK graviton coupling is larger due to its IR localisation.

\section{Diphoton resonance}

Suppose that  the resonance $X$ with spin $J$ decays into diphotons in cascade  through light intermediate particles  \cite{Knapen,photonjets,Dasgupta,axionmed}, namely, $X\rightarrow YY\rightarrow 2\gamma_{\rm jet}$, with $Y\rightarrow n\gamma\equiv \gamma_{\rm jet}$ with $n=2$ or 4, such that $n$ photons coming from the decay of each $Y$ are collimated and are considered as a singlet photon in the detector.
In this case, the resonance production cross section of particle $X$ with cascade decays is
\bea
\sigma(pp\rightarrow X\rightarrow 2\gamma_{\rm jet})=\frac{2J+1}{sM_X \Gamma_X}\, K_{gg} C_{gg}\Gamma(X\rightarrow gg) \Gamma(X\rightarrow YY) ({\rm BR}(Y\rightarrow \gamma_{\rm jet}))^2
\eea
where $C_{gg}$ is the gluon luminosity for the KK graviton production at $s=(13\,{\rm TeV})^2$ and the $K$-factor is given by $K_{gg}=1.4$ for a TeV-scale KK graviton.
Then, we assume that the $Y$ couplings to gauge bosons are small enough to evade the bounds on a light scalar at the LEP or LHC but they are large enough to decay within the ECAL. 
In our model, we choose the KK graviton as $X=G$ and the light radion as $Y=r$.

In this section, we first focus on the possibility of a light radion decaying into photons in cascade and and consider the bounds on the light radion in the parameter space where the collimated photons can mimick the diphoton resonance. 
Then, we continue to discuss the possibility of distinguishing the photon-jet contribution from the singlet photon contribution by converted photons as well as the angular distribution of photons.

\subsection{Photon-jets and bound on light radion}

%$1 (GeV)^{-1}= 0.1973\times10^{-13} cm$ \\
The hadronic decays of a light scalar with sub-GeV mass is the subject of much debate \cite{donoghue,bezrukov,volkas}. 
For $m_r<2 m_\pi\approx 280\,{\rm MeV}$, the radion decays only into a photon pair in the model with suppressed lepton couplings.
For $2m_\pi\approx 280\,{\rm MeV}<m_r<2m_K\approx 900\,{\rm MeV}$, the radion decays mostly into a pair of pions, so the branching fraction into a photon pair is negligible. For $2m_K<m_r\lesssim 1.5\,{\rm GeV}$, both $\pi\pi$ and  $KK$ decay modes are dominant. 
For $m_r\gtrsim 2.5\,{\rm GeV}$, the description of the radion decaying into a gluon pair is well justified.
However, there is a gap between $m_r=1.5-2.5\,{\rm GeV}$ for which neither the meson description nor the quarks/gluon description is correct, so there are large uncertainties in the hadronic decay rates \cite{bezrukov}.  In the following, we discuss the photon-jet contribution from the radion decay in each relevant case. 

For $m_r<280\,{\rm MeV}$, the radion decays only into a pair of photons, of course this assumes that the light leptons are localized towards the UV brane so that their wave-function overlap with the radion is quite suppressed.
Then, the decay length of the radion is estimated as
\begin{eqnarray}
L=\frac{1}{\Gamma_r}\cdot \frac{E_r}{m_r}= 1.5\, {\rm m}\bigg(\frac{0.28\,{\rm GeV}}{m_r}\bigg)^4\bigg(\frac{\Lambda/d_{\gamma\gamma}}{100\,{\rm TeV}} \bigg)^2.
\end{eqnarray}
So, depending on the  radion mass or $d_{\gamma\gamma}/\Lambda$, we can have a  long lived radion, which is bounded by the condition that the radion decays inside ECAL for the diphoton excess. 
On the other hand, for $m_r>1.5\,{\rm GeV}$ including the ambiguous region for the effective description, we include the radion decay into a gluon pair, leading to the decay length of the radion as 
\begin{eqnarray}
L=\frac{1}{\Gamma_r}\cdot \frac{E_r}{m_r}= 2.0\times 10^{-4}\, {\rm m}\bigg(\frac{1.5\,{\rm GeV}}{m_r}\bigg)^4\bigg(\frac{\Lambda/d_V}{100\,{\rm TeV}} \bigg)^2.
\end{eqnarray}
with $d_V\equiv d_{\gamma\gamma}=d_{gg}$.
Then, assuming that the gluon fusion production the KK graviton, we find that the production cross sections for direct diphotons and photon-jets coming from the decay of the KK graviton lead to the following conditions, respectively,
\be
\Gamma_G(gg)\Gamma_G(\gamma\gamma)= 1.24\times 10^{-5}{\rm GeV}^2 \Big(\frac{\sigma(pp\rightarrow \gamma\gamma)}{6\,{\rm fb}} \Big) \Big(\frac{\Gamma_G}{0.1\,{\rm GeV}}\Big)\left(\frac{2137}{C_{gg}}\right), 
\ee
\be
\Gamma_G(gg)\Gamma_G(rr)({\rm BR}(r\rightarrow\gamma\gamma))^2 =1.24\times 10^{-5}{\rm GeV}^2 \Big(\frac{\sigma(pp\rightarrow \gamma_{\rm jet}\gamma_{\rm jet})}{6\,{\rm fb}} \Big) \Big(\frac{\Gamma_G}{0.1\,{\rm GeV}}\Big)\left(\frac{2137}{C_{gg}}\right). 
\ee

\begin{figure}[t!]
  \begin{center}
   \includegraphics[height=0.5\textwidth]{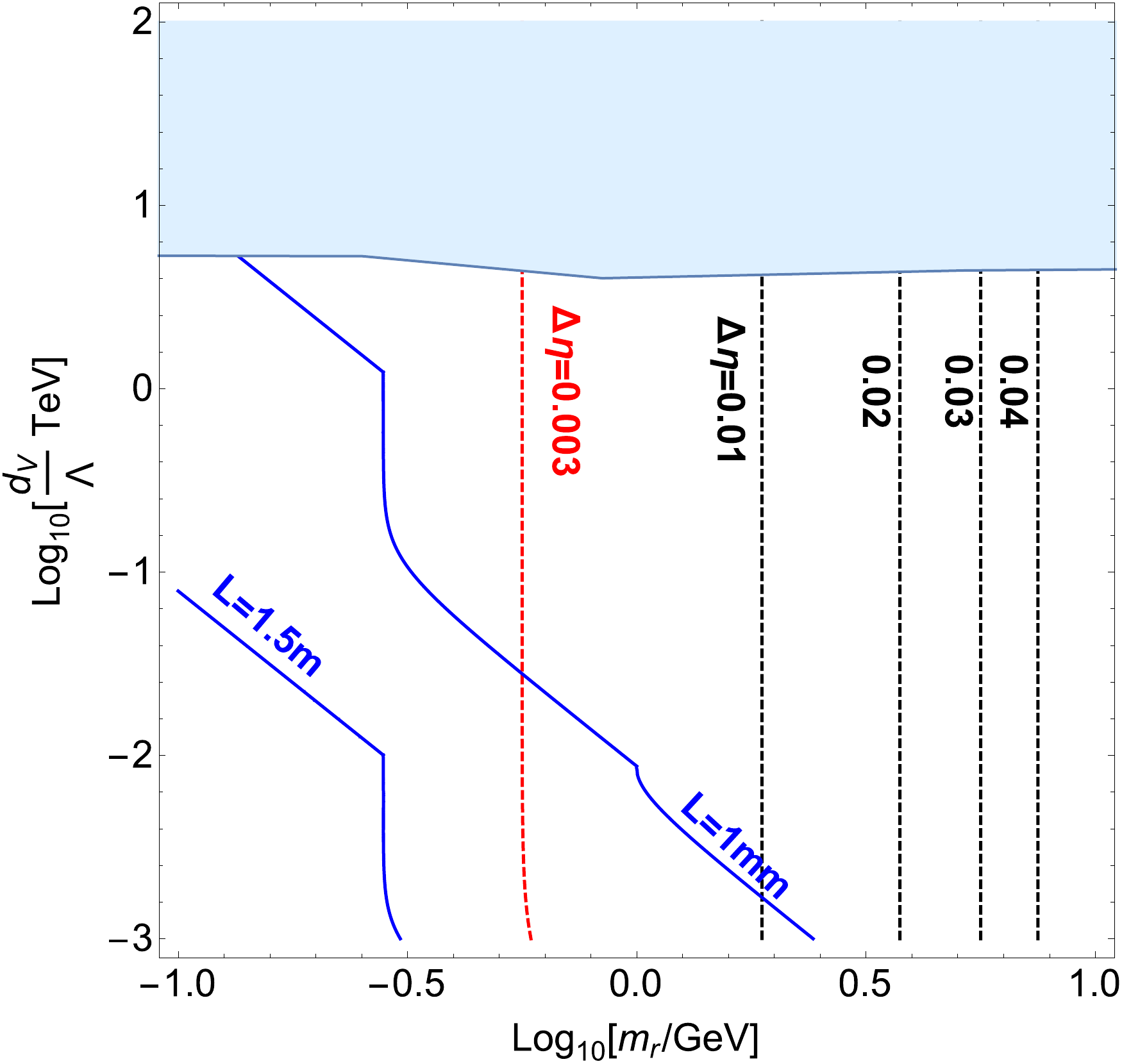}
   \end{center}
  \caption{ $\Delta\eta_{\gamma\gamma}$ between two collimated photons from the radion decay and the decay length of the radion in the parameter space, $m_r$ vs $d_V/\Lambda [{\rm TeV}^{-1}]$ with $d_V\equiv d_{\gamma\gamma}=d_{gg}$.  Current exclusion are adapted from the analyses in Refs.~\cite{Ken,Spanno}.}
  \label{eta}
\end{figure}

For $2m_\pi\approx 280\,{\rm MeV}<m_r<2m_K\approx 900\,{\rm MeV}$, the partial decay rate of the radion decaying into a pair of pions ($\pi^a\pi^a$ with $a=1,2,3$) is calculated in chiral perturbation theory as discussed in the previous section. 
Then, the ratio of decay rates of the pion and would-be gluon decay modes is given by $\Gamma_r(\pi^a\pi^a)/\Gamma_r(gg)=6-7.5$, depending on the radion mass.
Thus, for $d_{\gamma\gamma}=d_{gg}$, the radion decays mostly into a pair of pions. 
In this region, the decay length is estimated as follows,
\be
L=\frac{1}{\Gamma_r}\cdot \frac{E_r}{m_r}\simeq 8.2\times 10^{-4} \, {\rm m}\bigg(1+\frac{m^2_\pi}{m^2_r} \bigg)^{-2}\bigg(1-\frac{4m^2_\pi}{m^2_r}\bigg)^{-1/2}\bigg(\frac{0.5\,{\rm GeV}}{m_r}\bigg)^4\bigg(\frac{\Lambda/d_V}{100\,{\rm TeV}} \bigg)^2
\ee
with $d_V\equiv d_{\gamma\gamma}=d_{gg}$.
In this case, the diphoton events have a photon-jet contribution, produced from the decay of neutral pions in the cascade decay of the radion. Then, the production cross section for photon-jets coming from two step decays of the KK graviton leads to the condition,
\be
\Gamma_G(gg)\Gamma_G(rr)({\rm BR}(r\rightarrow\pi^0\pi^0))^2
=1.24\times 10^{-5}{\rm GeV}^2 \Big(\frac{\sigma(pp\rightarrow \gamma\gamma)}{6\,{\rm fb}} \Big) \Big(\frac{\Gamma_G}{0.1\,{\rm GeV}}\Big)\left(\frac{2137}{C_{gg}}\right).  \label{diphoton2}
\ee
For $m_r>2m_{\pi^\pm}$, the branching fraction of the radion into a pair of neutral pions is about ${\rm BR}(r\rightarrow \pi^0\pi^0)\simeq \frac{1}{3}$. But, for $2m_{\pi^0}< m_r < 2m_{\pi^\pm}$, the radion can decay almost fully into a pair of neutral pions. In the latter case, it is possible to have a large photon-jet contribution from four photons per each radion decay. 

Finally, for $2m_K<m_r<1.5\,{\rm GeV}$, the radion can also decay into a pair of kaons and/or a pair of $\eta$ mesons, on top of pions. The partial decay rates for these decays can be computed in a similar way to the pion decay modes, but with the inclusion of the $s$ quark in chiral perturbation theory. For instance, the decay rate into a pair of kaons is given by eq.~(\ref{pions}) with $b=9$ and $m_\pi$ being replaced by $m_K$. In this region, the decay rates of the radion into a photon-jet is further suppressed due to the presence of extra meson decay modes.

In Fig.~\ref{eta}, we show the separation of two collimated photons in pseudo-rapidity, $\Delta \eta_{\gamma\gamma}$, and the decay length $L$, in the parameter space of the effective coupling $d_V/\Lambda[{\rm TeV}^{-1}]$ and mass $m_r$ of radion.  In this case,  the pseudo-rapidity separation can be smaller than $0.003$ as required from the least cell size of the first layer in ATLAS ECAL. 
In the presence of the radion coupling to $\gamma\gamma$, we can also impose the LEP limit \cite{LEP,Ken} from $e^+e^-\rightarrow \gamma^* \rightarrow a\gamma$ with axion-like particle decaying by $a\rightarrow\gamma\gamma$.  If there is a $Z\gamma$ coupling to the radion, which is the case for $d_1\neq d_2$, a stronger limit from $e^+e^-\rightarrow Z\rightarrow a\gamma$ at the $Z$ pole in LEP \cite{Spanno} also applies in Fig.~\ref{eta}. But, in the case with $d_1=d_2$ which is our main focus in the later discussion, there is no bound from the $Z$ pole. 
On the other hand, in the region above the pion threshold with $m_r> 2m_\pi$, the branching fraction of $r\rightarrow \gamma\gamma$ gets suppressed so the LEP limit disappears in the figure.

\begin{figure}
  \begin{center}
       \includegraphics[height=0.5\textwidth]{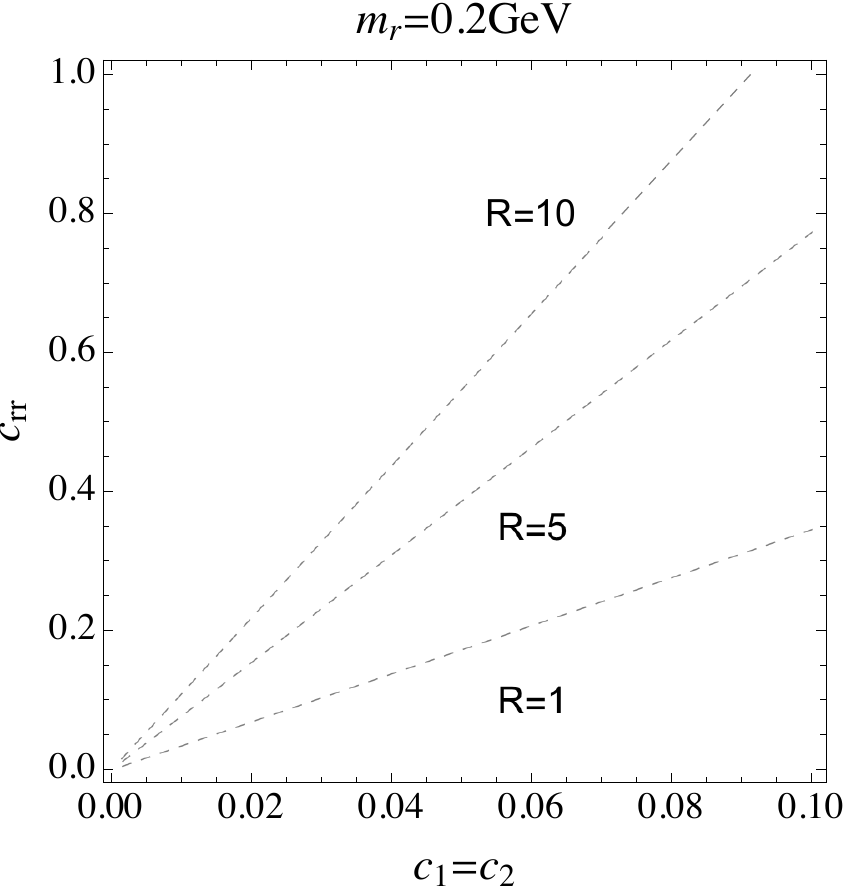}
   \end{center}
  \caption{Contour plots of $R$ values for photon-jet contribution in $c_1=c_2$ vs $c_{rr}$. We took the radion masses to $m_r=0.2\,{\rm GeV}$ while $c_H=c_f=0$.  }
  \label{couplings}
\end{figure}

Defining $R$ as the ratio of photon-jet and single photon contributions to the diphoton bump as follows, 
\be
R\equiv \frac{\Gamma_G(rr)({\rm BR}(r\rightarrow\gamma_{\rm jet}{\rm 's}))^2}{\Gamma_G(\gamma\gamma)},
\ee 
with $\gamma_{\rm jet}=2\gamma$ or $4\gamma$,
in Fig.~\ref{couplings}, we draw the contour plots with constant $R$ in the parameter space of $c_1=c_2$ vs $c_{rr}$ for $m_r=0.2\,{\rm GeV}$.

\subsection{Phenomenology of photon-jets}

In this section, we discuss the possibility  of distinguishing the photon-jet contribution in our model from the single photon contribution by converted photons.  
Moreover, we also show that the angular distributions of photons can be used to infer the information of direct or cascade decays of the KK graviton, in comparison to the direct decay of a spin-0 resonance.

\subsubsection{Converted photons}

A photon candidate is reconstructed from the energy deposited in the ECAL after different shape variables of energy profiles are imposed. For instance, the ATLAS  ECAL \cite{ATLAS-ECAL} consists of three layers. The first layer is segmented into high granularity strips with $\Delta\eta=0.003$-$0.006$ depending on $\eta$, with which we are able to make the $\gamma$-$\pi^0$ discrimination. The second layer of the ECAL has a granularity of 0.025$\times$0.025 in $\Delta\eta \times \Delta\phi$ (corresponding to one cell).  Generally if a bunch of highly collinear photons (``photon-jets") are contained in a very small cone with $\Delta R \lesssim 0.003$, the photon bunch could possibly pass the photon selection and be taken as a single photon\footnote{A dedicated simulation is needed to show the exact percentage of faked photons coming from photon-jets.}.

In our model, depending on the radion mass and couplings, the diphoton signal could come from two photons or one photon $+$ one photon-jet or two photon-jets. Even the case of two photon-jets could be divided into two categories: one category is that the photon-jet contains two photons coming from the direct decay of the radion and the other category is that the photon-jet contains four photons coming from the cascade decay of the radion into 2$\pi^0$. Therefore, it is very crucial to identify the contribution of photon-jets events from the diphoton events for extracting the model parameters.  

In addition to the energy deposition from the ECAL, a photon could be converted into electron and positron pair while traveling in the inner detector,  resulting in a photon candidate with tracks. Thus, the photon candidates with tracks are taken as ``converted" photons, and are otherwise taken as ``unconverted" photons. The photon conversion rate is well studied from experiment and it varies from 0.2 to 0.4 depending on $\eta$ and $p_T$ of the photon \cite{conversion}. Since photon-jets contain more than one photon, the conversion rate is expected to be larger than that of a single photon. Therefore, the number of ``converted" and ``unconverted" photons could provide information on the origin of the diphoton signal.

There have already been some studies on distinguishing the photon-jets from a single photon by using the number of ``converted" photons as well as the energy deposition in the ECAL \cite{Ellis, Knapen, Dasgupta}. In Ref.~\cite{Ellis}, it was shown that the substructure variables could help to discriminate photon-jets from single photon if the separation of photons inside the photon-jets is large enough ($\Delta R\sim 0.01$). A more dedicated analysis on the diphoton resonance is performed in Ref.~\cite{Dasgupta} from the counting of the number of ``converted" photons. They provided the required number of events to distinguish between the case where the diphoton reaonance is solely from photon-jets, one genuine photon per photon-jet, and the case where the diphoton resonance is from two single photons. In Ref.~\cite{Knapen}, the energy deposition of two photons is utilized.  All the previous methods must be confirmed by experimentalists and especially, more information about the number of ``converted"  photons  is necessary before any conclusion is drawn.

\subsubsection{Angular distributions}

When two photons (or a fake photon, in our case, a photon-jet) is produced from the decay of a heavy resonance ($X$) at the LHC, we can determine the properties of $X$ only by the phase space variable, $\cos\theta^*$, the scattering angle of a photon relative to the beam axis at the rest frame of $X$. In the case where $X$ is a KK graviton $(G)$, the differential cross section with respect to $\cos\theta^*$ depends on  the initial parton states for the production of $G$ \cite{Character}. In the case of the diphoton resonance, the gluon initial states $(gg \to X)$ are favoured, as the luminosity increase in PDF at $13\,{\rm TeV}$ makes the diphoton resonance compatible with the $8\,{\rm TeV}$ data. 
The corresponding angular distributions of the two-photon channel would be
\begin{equation}
\frac{\textrm{d} \hat \sigma}{\textrm{d} \cos\theta^*} \propto 
\begin{cases}
1+6\cos^2\theta^*+\cos^4\theta^* & \text{in }(gg \to G \to \gamma \gamma) ,\\
1-2\cos^2\theta^*+\cos^4\theta^* & \text{in }(gg \to G \to r\, r \to 2\gamma_{\textrm{jet}}) , \\
1& \text{in }(gg \to S \to \gamma \gamma) ,\\
\end{cases}
\end{equation}
with $X$ being a spin-2 KK graviton in the first two lines and, for comparison, a scalar particle $(S)$ in the last line. In order to make a precise determination of the particles properties using the angular variable, we require more statistics than for the discovery \cite{Fabbrichesi:2016jlo}.
But the different kinematical behavior also affects the efficiencies of the analysis cuts, the rapidity coverage of the detector geometry, for example. 
From the simple relation between $\cos\theta^*$ and the rapidity $\eta$, we rewrite the angular distributions in the following way,
\begin{equation}
\frac{\textrm{d} \hat \sigma}{\textrm{d} \eta^*} \propto 
\begin{cases}
\cosh(4\eta^*) \sech^6 \eta^* & \text{in }(gg \to G \to \gamma \gamma) ,\\
\sech^6 \eta^* & \text{in }(gg \to G \to r\, r \to 2\gamma_{\textrm{jet}}) , \\
\sech^2\eta^*& \text{in }(gg \to S \to \gamma \gamma).\\
\end{cases}
\end{equation}
Here we use $\eta^*$ to denote the rest frame, or center of mass frame, of $X$.

%%%%%%%%%%%%%%%%%%%%%%%%%%%FIGURE%%%%%%%%%%%%%%%%%%%%%%%%%
%%%%%%%%%%%%%%%%%%%%%%%%%%%FIGURE%%%%%%%%%%%%%%%%%%%%%%%%%
\begin{figure}
  \begin{center}
   \includegraphics[height=0.3\textwidth]{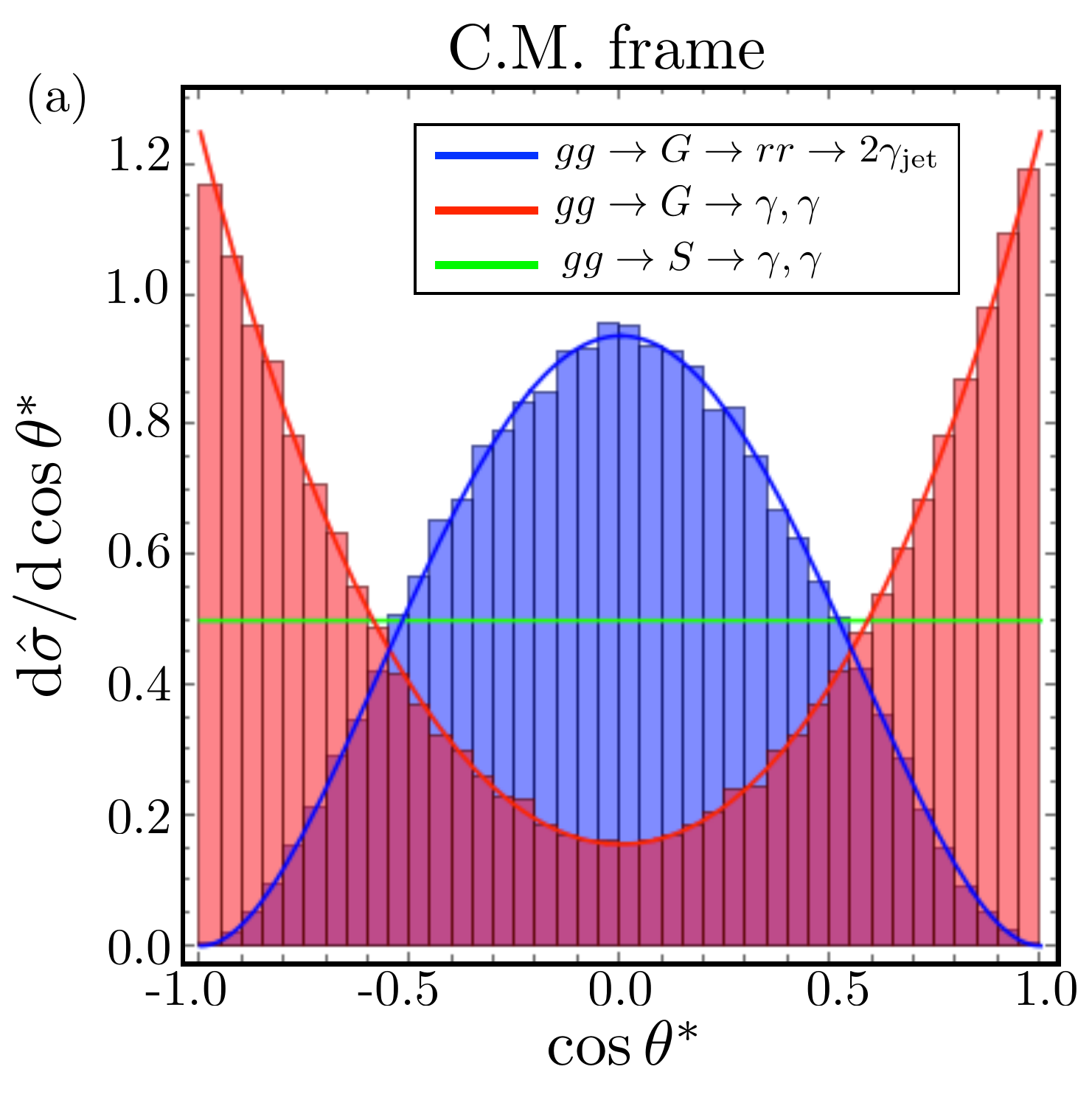}
      \includegraphics[height=0.3\textwidth]{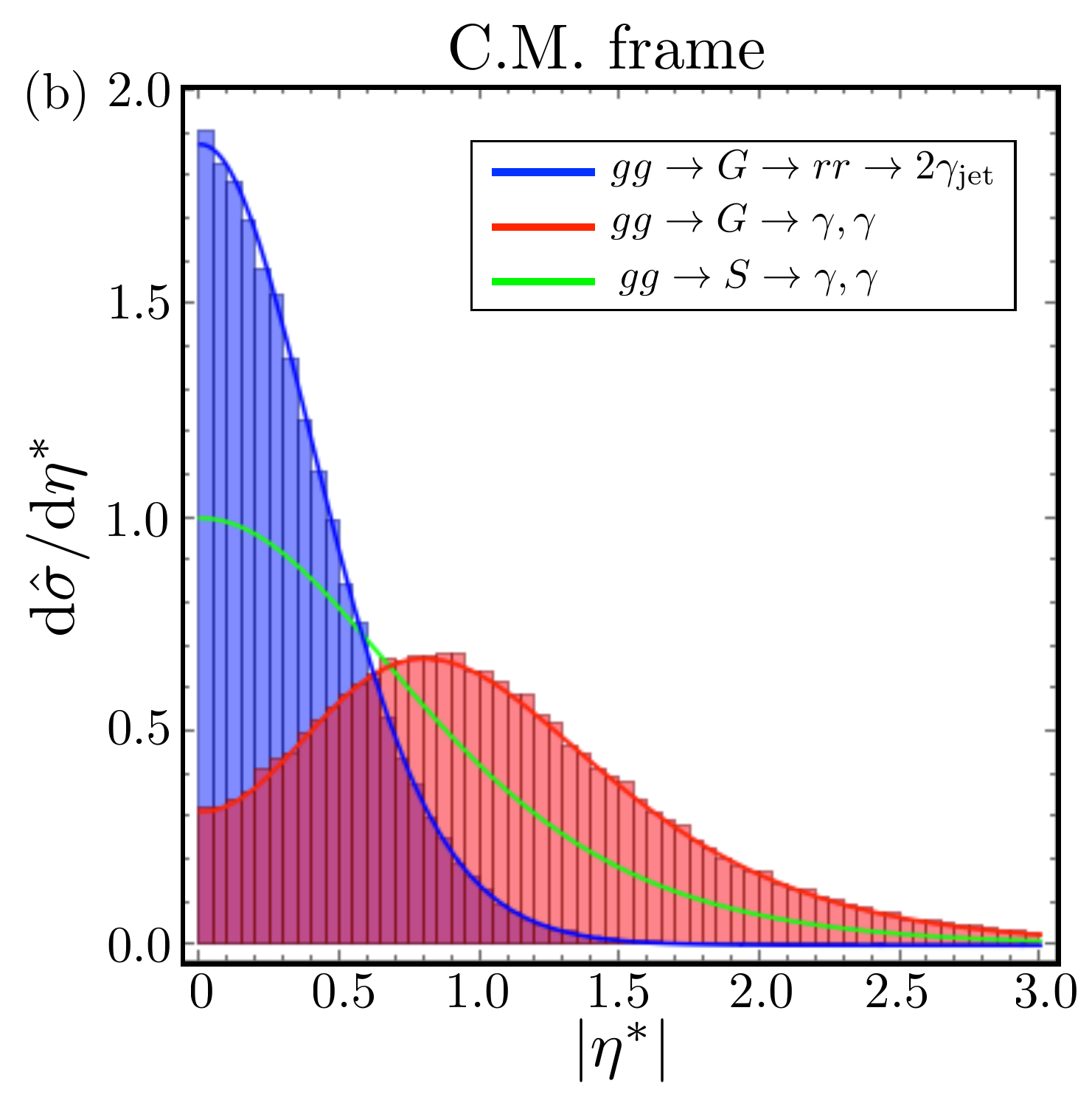}
          \includegraphics[height=0.29\textwidth]{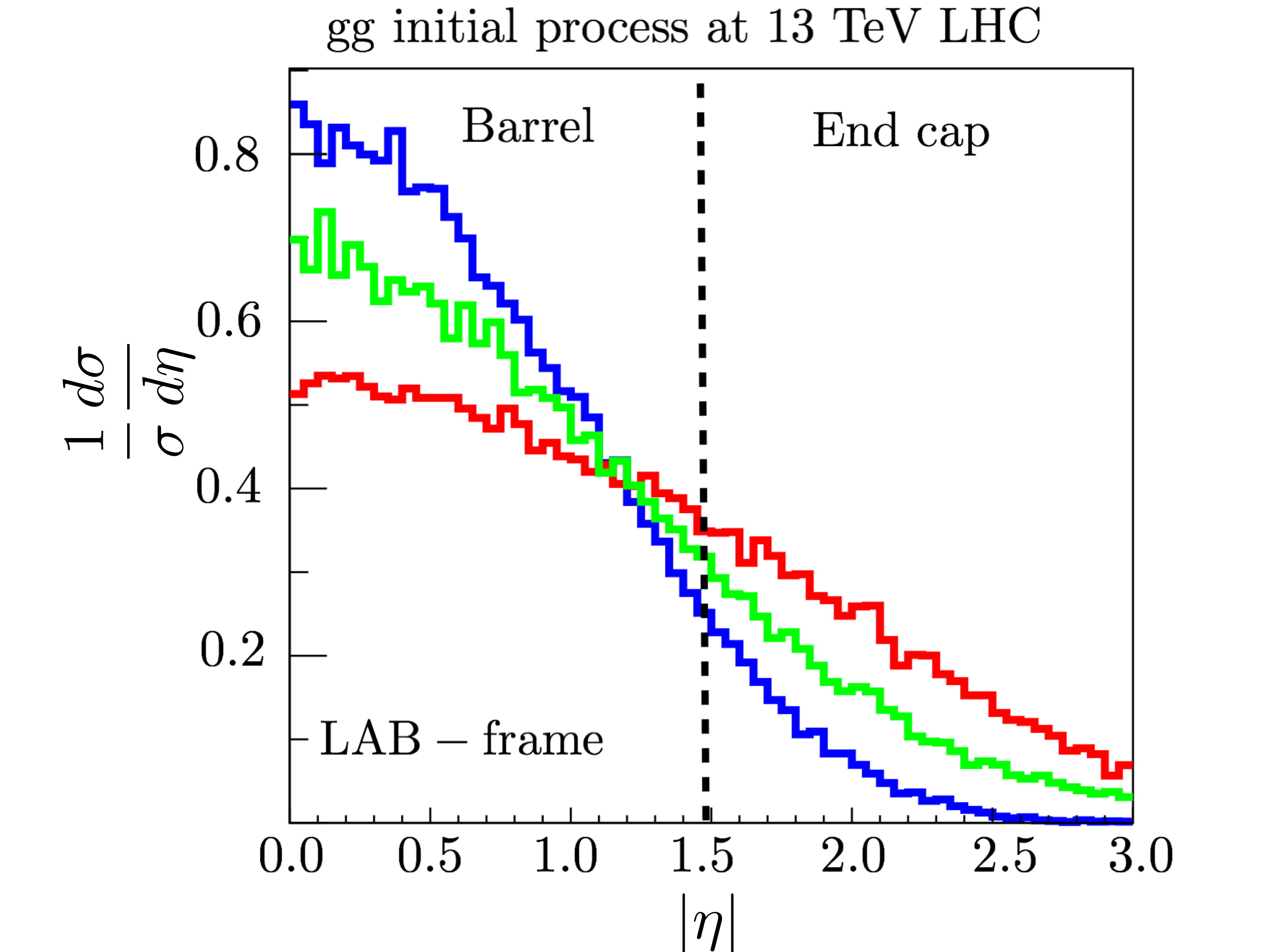}
   \end{center}
  \caption{ (a) Distributions of photons for the scattering angle $\cos\theta^*$, (b) pseudo-rapidity $\eta^*$ in the CM frame of the KK graviton and (c) $\eta$ at the LAB frame. 
  Red lines correspond to $G\rightarrow \gamma\gamma$ while blue lines correspond to $G\rightarrow rr\rightarrow 2\gamma_{\rm jet}$ with $\gamma_{\rm jet}$ as a photon-jet.
   For comparison, we add the case for a spin-0 resonance with $S\rightarrow \gamma\gamma$ in green lines. Here, we took $m_G=1\,{\rm TeV}$, $m_r=0.2\,{\rm GeV}$, $\Gamma_G=10\,{\rm GeV}$ and $\Gamma_r=6\times 10^{-10}\, {\rm GeV}$. }
  \label{eta2}
\end{figure}
%%%%%%%%%%%%%%%%%%%%%%%%%%%FIGURE%%%%%%%%%%%%%%%%%%%%%%%%%
%%%%%%%%%%%%%%%%%%%%%%%%%%%FIGURE%%%%%%%%%%%%%%%%%%%%%%%%%

In Fig.~\ref{eta2}, we have shown the distributions of photons as a function of the scattering angle $\cos\theta^*$ in (a) and the pseudo-rapidity $\eta^*$ in (b) in the Center of Mass (CM) frame of the KK graviton.  We took the KK graviton to $1 \, {\rm TeV}$ and the radion mass to $0.2 \,{\rm GeV}$. The decay width of the radion is taken to $6\times 10^{-10}\, {\rm GeV}$ for $d_V/\Lambda=(1\,{\rm TeV})^{-1}$ where the width of the KK graviton is assumed to be $10 \,{\rm GeV}$ in the plot.
After the PDF convolution, the differential distributions of $\eta$ in the LAB frame can be obtained from the above $\eta^*$ distributions.
To see the effect of gluon PDFs, we performed a parton-level Monte Carlo simulation as shown in Fig.~(\ref{eta2}) with Madgraph \cite{Alwall:2014hca}. 
As a result, in the case where a light radion induces fake photons, many of the events would be in the central(Barrel) region of the detector.
In the case with the direct decay of a KK graviton, more than $30\%$ events would be in the end cap region as in Fig.~\ref{eta2}-(c). In our case with photon-jets, however, only about $15\%$ events would be in the End-cap region. The results can be compared to
the case of a scalar particle where $25\%$ events would be in the End-cap region.
Therefore, the optimized search for the cascade decays in our model ($G\rightarrow rr \rightarrow 2 \gamma_{\rm jet}$) would be closer to the case for spin-0 resonance rather than the case for spin-2 resonance with direct decays, due to the population of events in the central region of the detector.

\section{KK graviton and radion as mediators of dark matter}

Suppose that dark matter is localized toward the IR brane. In this case, both the KK graviton and the radion can have sizeable couplings to dark matter so that they can mediate between dark matter and the SM particles \cite{gmdm,tensor}. Given that the radion is sub-GeV, the radion mediated channels give dominant contributions to the total annihilation cross section of dark matter, away from the resonance with KK graviton exchanges. 

The effective interactions of KK graviton and radion to dark matter are given by
\bea
{\cal L}_{\rm KK}
=-\frac{1}{\Lambda}G^{\mu\nu}T^{\rm DM}_{\mu\nu}+\frac{1}{\sqrt{6}\Lambda}\, r\, T^{\rm DM},
\eea
where the energy-momentum tensor for dark matter, depending on the spin ($0,1/2,1$) of dark matter, is given by
\bea
T^{\rm S}_{\mu\nu}&=&c_S\bigg[ \partial_\mu S \partial_\nu S-\frac{1}{2}g_{\mu\nu}\partial^\rho S \partial_\rho S+\frac{1}{2}g_{\mu\nu}  m^2_S S^2\bigg],\\
T^{\rm F}_{\mu\nu}&=& c_\chi \bigg[\frac{i}{4}{\bar\chi}(\gamma_\mu\partial_\nu+\gamma_\nu\partial_\mu)\chi-\frac{i}{4} (\partial_\mu{\bar\chi}\gamma_\nu+\partial_\nu{\bar\chi}\gamma_\mu)\chi-g_{\mu\nu}(i {\bar\chi}\gamma^\mu\partial_\mu\chi- m_\chi {\bar\chi}\chi) \bigg]
\nonumber \\
&&+\frac{i}{2}g_{\mu\nu}\partial^\rho({\bar\chi}\gamma_\rho\chi)\bigg],  \nonumber \\
T^{\rm V}_{\mu\nu}&=&
c_X\bigg[ \frac{1}{4}g_{\mu\nu} X^{\lambda\rho} X_{\lambda\rho}+X_{\mu\lambda}X^\lambda\,_{\nu}+m^2_X\Big(X_{\mu} X_{\nu}-\frac{1}{2}g_{\mu\nu} X^\lambda  X_{\lambda}\Big)\bigg].
\eea
Here we note that the couplings of the KK graviton and the radion to dark matter can be different, so $c_{S,\chi,X}$ should be replaced by $c^G_{S,\chi,X}$ and $c^r_{S,\chi,X}$ for the KK graviton and radion couplings, respectively. 
In the presence of the couplings between the KK graviton and dark matter, the KK graviton could decay invisibly into a pair of dark matter particles, if kinematically allowed. The corresponding partial decay rates of the invisible decays are
\bea
\Gamma(SS)&=&  \frac{(c^G_S)^2 m^3_G}{960 \pi \Lambda^2} \Big(1-\frac{4m^2_S}{m^2_G}\Big)^\frac{5}{2}, \\
\Gamma(\chi{\bar\chi})&=& \frac{ (c^G_\chi)^2 m^3_G}{160 \pi \Lambda^2}
 \left(1-\frac{4m^2_\chi}{m^2_G} \right)^\frac{3}{2} \left(1+\frac{8}{3} \frac{m^2_\chi}{m^2_G}\right), \\
 \Gamma(XX)&=& \frac{ (c^G_X)^2 m^3_G}{960\pi \Lambda^2}\Big(1- \frac{4m^2_X}{m^2_G}\Big)^\frac{1}{2}
\Big(13+\frac{56m^2_X}{m^2_G}+\frac{48m^4_X}{m^4_G}\Big).
\eea

\begin{figure}
  \begin{center}
    \includegraphics[height=0.50\textwidth]{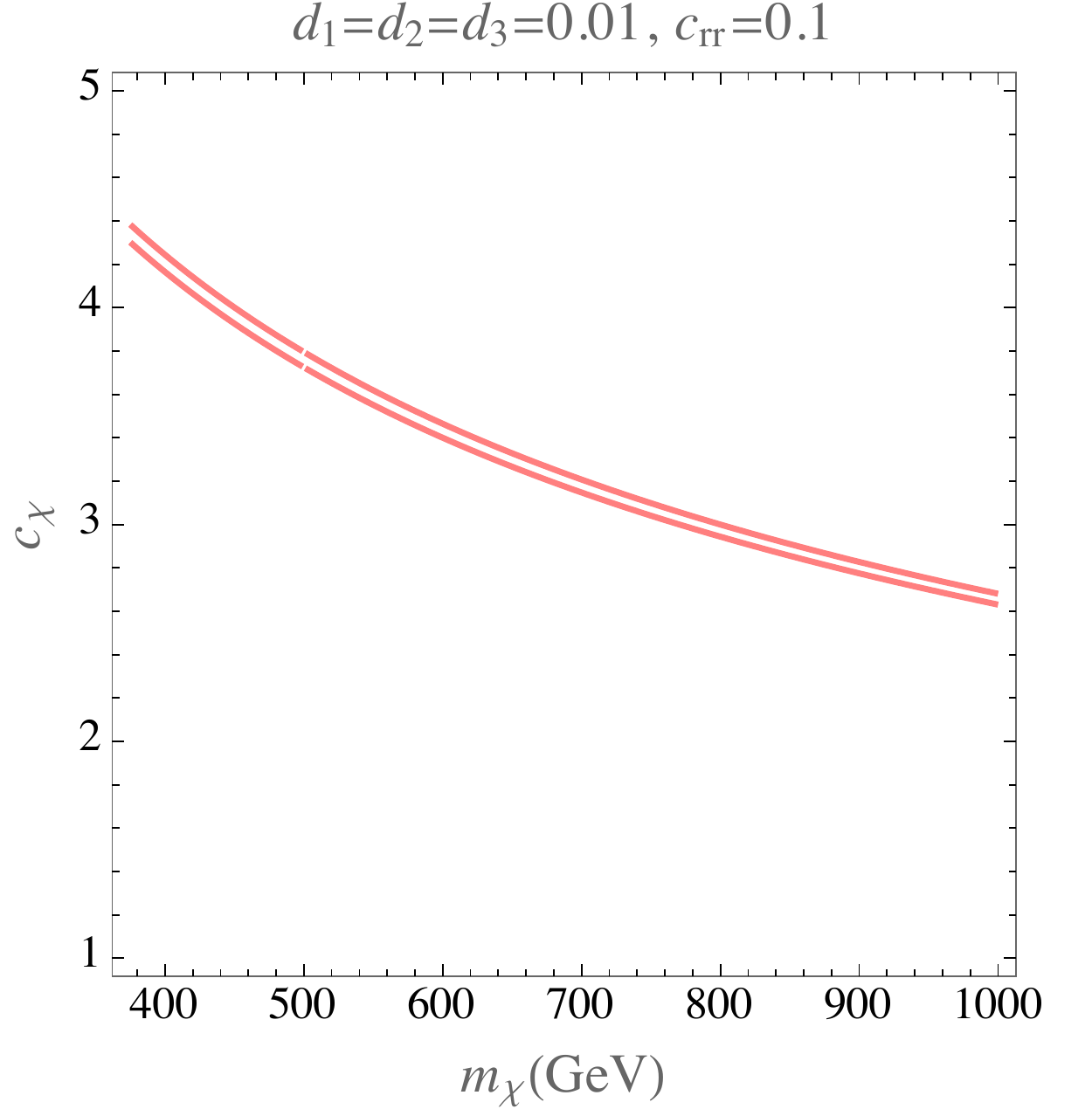}
     \end{center}
  \caption{Parameter space for dark matter mass ($m_\chi$) and radion coupling to dark matter ($c^r_\chi$), satisfying the relic density given by Planck within $3\sigma$.  We have taken $\Lambda=1\,{\rm TeV}$, $m_G=1\,{\rm TeV}$, $m_r=0.4\,{\rm GeV}$,  $d_1=d_2=d_3=0.01$ and $c_{rr}=0.1$. \label{relic}}
  \label{relic}
\end{figure}

Due to dilatation symmetry, the non-derivative couplings of radion are fixed beyond the linear order \cite{gmdm} as 
\bea
{\cal L}_{\rm non-deriv}&=&2\bigg(\frac{r}{\sqrt{6}\Lambda}
-\frac{r^2}{3\Lambda^2}\bigg)c^r_S m^2_S S^2  + \frac{r}{\sqrt{6}\Lambda} c^r_\chi m_{\chi} \bar{\chi} \chi \nonumber \\
&&-\bigg(\frac{r}{\sqrt{6}\Lambda}-\frac{r^2}{6\Lambda^2}\bigg)c^r_X m^2_X X_\mu X^\mu.
\eea

For illustration purposes, in the following discussion we focus on Dirac fermion dark matter with a radion mediator and consider the radion couplings to transverse gauge bosons and a KK graviton only, i.e. with $d_H=d_V=d_f=0$. 
Then, the partial annihilation cross sections into a pair of SM gauge bosons are given \cite{axionmed} by 
\bea
(\sigma v_{\rm rel})_{gg}&=& \frac{4(c^r_\chi)^2 d^2_{gg}}{3\pi \Lambda^4}\,\frac{m^6_\chi}{(4m^2_\chi-m^2_r)^2+\Gamma^2_r m^2_r}\,v^2_{\rm rel}, \\
(\sigma v_{\rm rel})_{\gamma\gamma}&=& \frac{(c^r_\chi)^2 d^2_{\gamma\gamma}}{6\pi\Lambda^4}\,\frac{m^6_\chi}{(4m^2_\chi-m^2_r)^2+\Gamma^2_r m^2_r}\,v^2_{\rm rel}, \\
(\sigma v_{\rm rel})_{Z\gamma} &=&  
 \frac{(c^r_\chi)^2 d^2_{Z\gamma}}{12\pi\Lambda^4}\,\frac{m^6_\chi}{(4m^2_\chi-m^2_r)^2+\Gamma^2_r m^2_r}\,\left( 1-\frac{m^2_Z}{4m^2_\chi} \right)^3 v^2_{\rm rel}, \\
 ( \sigma v_{\rm rel})_{ZZ} &=&  \frac{(c^r_\chi)^2 d^2_{ZZ}}{6\pi\Lambda^4} \,\frac{m^6_\chi v^2_{\rm rel}}{(4m^2_\chi-m^2_r)^2+\Gamma^2_r m^2_r}\,\Big(1-\frac{m^2_Z}{m^2_\chi}+\frac{3m^4_Z}{8m^4_\chi}\Big)\left(1-\frac{m^2_Z}{m^2_\chi} \right)^{1/2} , \\
 ( \sigma v_{\rm rel})_{WW}&=&  \frac{(c^r_\chi)^2 d^2_{WW}}{12\pi\Lambda^4}\,\frac{m^6_\chi v^2_{\rm rel}}{(4m^2_\chi-m^2_r)^2+\Gamma^2_r m^2_r}\, \Big(1-\frac{m^2_W}{m^2_\chi}+\frac{3m^4_W}{8m^4_\chi}\Big)\left(1-\frac{m^2_W}{m^2_\chi} \right)^{1/2}
 \eea
where $d_{\gamma\gamma}=s_{\theta}^2d_2+c_{\theta}^2d_1$, $d_{ZZ}=c_{\theta}^2d_2+s_{\theta}^2d_1$, $d_{Z\gamma}=s_{\theta}c_{\theta}(d_2-d_1)$, $d_{gg}=d_3$, $d_{WW}=2d_2$,
We note that all the gauge boson channels are $p$-wave suppressed.   
Furthermore, for $m_\chi>m_r$, dark matter can annihilate into a pair of radions.  In the limit of non-relativistic dark matter, the corresponding annihilation cross section for the $rr$ channel \cite{axionmed} becomes  
\bea
(\sigma v_{\rm rel})_{rr}=  \frac{(c^r_\chi)^4}{288\pi\Lambda^4} \frac{ m^6_\chi(2(m^2_r-2m^2_\chi)^2+m^4_\chi)}{3(m^2_r-2m^2_\chi)^4} \Big(1-\frac{m^2_r}{m^2_\chi}\Big)^{1/2}\, v^2_{\rm rel}\,.
\eea
Thus, the $rr$ channel turns out to be also $p$-wave suppressed, but it can contribute to the thermal cross section at freeze-out. If $m_\chi>(m_G+m_r)/2$, the $rG$ channel would be open too and the corresponding annihilation cross section is given by
\bea
(\sigma v_{\rm rel})_{Gr}= \frac{(c^r_\chi)^2 c^2_{rr}}{9\pi\Lambda^4}\,\frac{m^{10}_\chi v^2_{\rm rel}}{((4m^2_\chi-m^2_r)^2+\Gamma^2_r m^2_r)m^4_G}\, \bigg(1-\frac{(m_G-m_r)^2}{4m^2_\chi} \bigg)^{5/2}   \bigg(1-\frac{(m_G+m_r)^2}{4m^2_\chi} \bigg)^{5/2}.   
\eea
We note that the KK graviton mediator also contributes to the annihilation of the above fermion dark matter too, becoming important near the resonance region. But, we ignore the contributions from the KK graviton, focusing on the region outside the resonance.

In Fig.~\ref{relic}, we show the parameter space for the dark matter mass and coupling to the radion, considering the bound on the relic density from Planck within $3\sigma$.  We took the radion couplings to the SM gauge bosons and to the KK graviton such that the photon-jet mode is dominant over the direct photon mode.  In this case, a sizable radion coupling to dark matter can accommodate a correct relic density.  
For $m_\chi<m_G/2$, there would be an invisible decay of the KK graviton into a pair of dark matter. If the KK coupling is of a similar order to the radion counterpart (namely, $c^G_\chi\sim c^r_\chi$), the invisible decay would be so large such that  the branching fractions of KK graviton decaying into $\gamma\gamma$  or $rr$ are suppressed. Thus, we took $m_\chi>m_G/2$ in Fig.~\ref{relic}. But, if the KK graviton coupling to dark matter gets smaller, the invisible decay rate of the KK graviton could be at the level of detectability in the near future at the LHC\cite{tensor}.

\section{Holographic composite Higgs and LHC bounds}

In the holographic composite Higgs scenario \cite{CompHiggs}, the Higgs field arises as $A_5$ components of a bulk gauge field, which are localised in the IR ($a_h=-1$), the right-handed top quark is required to be localised in the IR also ($a_{tR}=-1$), and the $(t_L,b_L)$ doublet is required to be delocalised ($a_q\sim0$).  This model is much different than that considered previously, and the experimental bounds from Run 1 and Run 2 can be more stringent due to the large coupling of the KK graviton to the Higgs.  However more stringent bounds arise between the KK graviton and the transverse gauge fields.

\begin{figure}
  \begin{center}
    \includegraphics[height=0.40\textwidth]{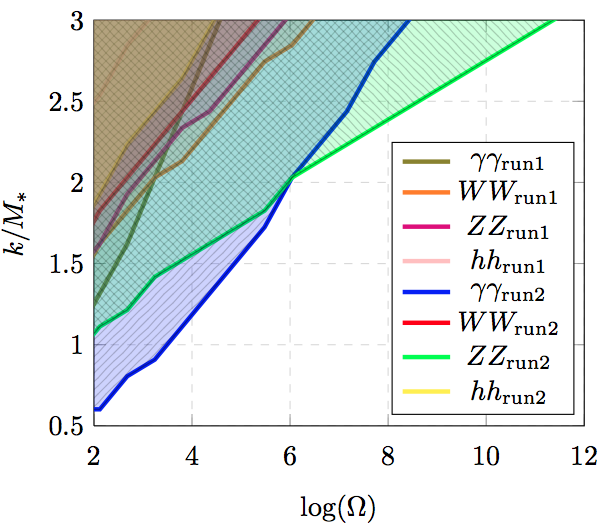}
    \includegraphics[height=0.40\textwidth]{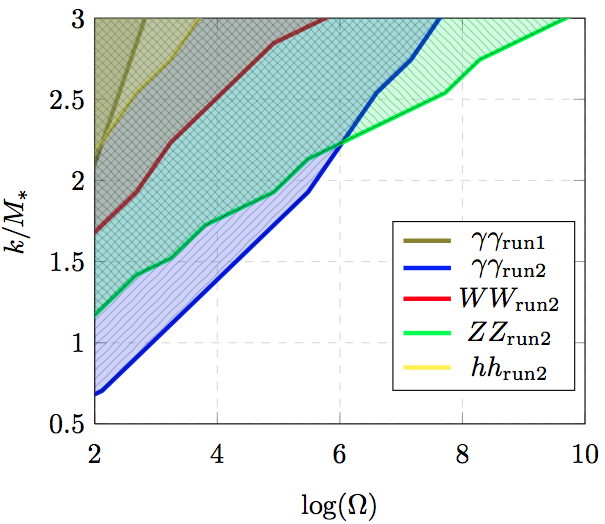}
   \end{center}
  \caption{Experimental bounds from run 1 and run 2 on the KK graviton parameters in the composite-Higgs model  \cite{Dillon:2016fgw}.  We assumed that the KK graviton is produced via gluon fusion then decaying in various channels.  We have used $m_G=1$ and 2 TeV, respectively, and taken $M_{ KK}=1 \,{\rm TeV}$ in each case. }
  \label{XBRs_noBKT}
\end{figure}

In Eq. \ref{BFradion}, we see that by lowering $kR\equiv \text{log}(\Omega)$ the photon-jet contribution to the di-photon signal becomes smaller.
At the same time this has the effect of increasing the KK graviton couplings to transverse gauge fields, and thus the experimental bounds from Run 1 and 2 become much more relevant.
In \cite{Dillon:2016fgw} the authors studied the phenomenology of KK gravitons while varying $\text{log}(\Omega)$, and they show that for $\text{log}({\Omega})\lesssim12$ there are exclusion bounds on the parameter $k/M_{\star}$, mostly arising from the di-photon and $ZZ$ final states.  
In Fig. \ref{XBRs_noBKT} we show the experimental bounds on the $k/M_{\star}$ and $\text{log}({\Omega})$ variables for $M_{KK}=1$ TeV and $m_G=1$ and $2$ TeV by using the results in  \cite{Dillon:2016fgw}.
We see that the di-photon bound is stronger than that from the ZZ final state for $\text{log}(\Omega)\lesssim6$.
In this region the photon-jet contribution to the total di-photon signal is less than $\sim14\%$, and for larger values of $\text{log}(\Omega)\lesssim6$ this increases.
As shown in Fig. \ref{BKTgaugeCouplingRatios}, the couplings of the KK graviton to photons can be naturally suppressed due to the effects of the brane kinetic term for the hypercharge gauge field.
The effects of this brane kinetic term on the radion couplings are minimal and thus the photon-jet signal would remain approximately constant, thus this could be the dominant contribution in the di-photon channel.

\section{Conclusions}

We have presented a theoretical framework to study the diphoton resonance as a manifestation of the interplay between direct decays to photons or photon-jets, originating from the sequential decay of the heavy resonance to a pair of new light resonances. We assumed the new resonances are spin-two and -zero particles (decaying in that order) motivated by theories of warped gravity at the TeV scale, and explored the phenomenology in the model-independent fashion. 

We showed that a large wavefunction overlap between the KK graviton and the radion in the warped extra dimension allows for a large gravity self-coupling between them in 4D effective theory, making the KK graviton decay largely into a pair of radions. As a light radion can decay mostly into a pair of photons, the resulting photon-jets from the decay of the KK graviton can mimick the diphoton resonance due to the limited $\eta$ resolution of single photons in the ECAL.  We provided an explicit example of the photon-jet scenario in the extra dimension, being consistent with current limits from direct searches and electroweak tests. 
Although we focused on the light radion in our paper, it would be also worthwhile to discuss in a future work the phenomenology of cascade decays of the KK graviton in a pair of heavy radions, each of which can decay into $WW, ZZ$, heavy fermions, etc.

We found that angular distributions of the two photon-jets could be a handle to distinguish between direct photon and photon-jet situations in our model. For instance, the direct photons are more directed towards the End-cap region while the photon-jets are more directed towards the central(Barrel) region.  Furthermore, photon-jets in our model turn out to be more central than single photons obtained from the direct decay of a spin-0 resonance.  

New resonances such as a KK graviton and a radion can play the role of the mediator of dark matter and could be seen from the invisible decay into a pair of dark matter plus mono-jet. We have illustrated that a sizable coupling of the radion to dark matter is necessary for weak-scale dark matter.

\section*{Acknowledgments}
We would like to thank Veronica Sanz for helpful discussions during the project. 
The work of CH is supported by the World Premier International Research Center Initiative (WPI Initiative), MEXT, Japan. 
The work of HML is supported in part by Basic Science Research Program through the National Research Foundation of Korea (NRF) funded by the Ministry of Education, Science and Technology (NRF-2016R1A2B4008759). 
MP is supported by IBS under the project code, IBS-R018-D1. 

\def\theequation{A.\arabic{equation}}

\setcounter{equation}{0}

\vskip0.8cm
\noindent
{\Large \bf Appendix:  Couplings between radion and KK graviton }
\vskip0.4cm
\noindent

In this appendix, we provide the details for the triple self-couplings between the radion and the KK graviton.  The couplings between these two fields come from the gravity self-interactions in 5D. 
For the warped metric, 
\be
ds^2= w^2(y){\cal G}_{\mu\nu} dx^\mu dx^\nu +G_{55}dy^2 \equiv G_{MN} dx^M dx^N
\ee
with ${\cal G}_{\mu\nu}=e^{-2{\hat r}}{\hat g}_{\mu\nu}$ and $G_{55}=-(1+2{\hat r})^2$, the 4D components of the Einstein tensor are 
\bea
R_{\mu\nu}({\cal G})= R_{\mu\nu}({\hat g})+ 2(\partial_\mu {\hat r}\partial_\nu {\hat r} +\partial_\mu \partial_\nu {\hat r}) +\eta_{\mu\nu} (\Box {\hat r}-2(\partial {\hat r})^2).
\eea
In linearized gravity with ${\hat g}_{\mu\nu}=\eta_{\mu\nu}+{\hat G}_{\mu\nu}$, we get
\bea
R_{\mu\nu}({\hat g})&=& \partial_\mu\partial_\nu {\hat G}^{\mu\nu}-\Box {\hat G} +\frac{1}{2} {\hat G}^{\mu\nu}\Big(\Box {\hat G}_{\mu\nu}+\partial_\mu\partial_\nu {\hat G}-\partial_\lambda\partial_\mu {\hat G}^\lambda_\nu-\partial_\lambda\partial_\nu {\hat G}^\lambda_\mu  \Big)+\cdots, \\
 \sqrt{-{\hat g}}&=&  1+\frac{1}{2}{\hat G}-\frac{1}{4}\Big({\hat G}^{\mu\nu} {\hat G}_{\mu\nu}-\frac{1}{2} {\hat G}^2\Big)+\cdots
\eea
where ${\hat G}\equiv \eta^{\mu\nu}{\hat G}_{\mu\nu}$ and $\cdots$ are higher order terms for ${\hat G}_{\mu\nu}$.
Then, the effective Lagrangian for the KK graviton and the radion is
\bea
{\cal L}_{\rm eff}
&=& \int dy \,w^2 (1+ 2{\hat r}) e^{-4{\hat r}} \sqrt{-{\hat g}}\bigg[e^{2{\hat r}} (\eta^{\mu\nu}-{\hat G}^{\mu\nu}) R_{\mu\nu}({\cal G})+G^{55} R_{55}(G) \bigg]  \\
&=&  \int dy \,w^2 \bigg[ -\frac{1}{2}{\hat G}\Box {\hat G}+ \frac{1}{2}{\hat G}^{\mu\nu}(\Box G_{\mu\nu}+\partial_\mu\partial_\nu {\hat G}-\partial_\lambda\partial_\mu {\hat G}^\lambda_\nu-\partial_\lambda\partial_\nu {\hat G}^\lambda_\mu  \Big)-6(\partial {\hat r})^2 \nonumber \\
&& -2{\hat G}^{\mu\nu}\partial_\mu{\hat r}\partial_\nu {\hat r} - 3{\hat G}(\partial {\hat r})^2 -2{\hat r}^2(\partial_\mu\partial_\nu {\hat  G}^{\mu\nu}-\Box {\hat G}) +\frac{3}{2} ({\hat G}^{\mu\nu}{\hat G}_{\mu\nu}-\frac{1}{2}{\hat G}^2) \Box {\hat r}+ \cdots \bigg]. \nonumber
\eea
Here, we note that the radion-KK graviton mixing terms are eliminated by the field definition of the graviton \cite{giudice}.
As a consequence, the triple couplings between KK graviton and radion are 
\bea
{\cal L}_{rr}&=&-2
\int dy \,w^2 \bigg[ {\hat G}^{\mu\nu}\partial_\mu{\hat r}\partial_\nu {\hat r} + \frac{3}{2}{\hat G}(\partial {\hat r})^2 +{\hat r}^2(\partial_\mu\partial_\nu {\hat  G}^{\mu\nu}-\Box {\hat G}) \bigg], \\
{\cal L}_{GG}&=&\frac{3}{2} \int dy \,w^2 \Big({\hat G}^{\mu\nu}{\hat G}_{\mu\nu}-\frac{1}{2}{\hat G}^2\Big) \Box {\hat r}.
\eea
Then, by inserting the bulk profiles for the KK graviton and the radion from (\ref{profiles}), the above triple effective interactions become
\bea
{\cal L}_{rr}&=&-\frac{c_{rr}}{\Lambda} \bigg[ {G}^{\mu\nu}\partial_\mu{ r}\partial_\nu { r} + \frac{3}{2}{G}(\partial { r})^2 +{r}^2(\partial_\mu\partial_\nu {G}^{\mu\nu}-\Box { G}) \bigg], \\
{\cal L}_{GG}&=&\frac{c_{GG}}{\Lambda} \Big(G^{\mu\nu}{G}_{\mu\nu}-\frac{1}{2}{G}^2\Big) \Box { r}
\eea 
where
\bea
c_{rr}&=&\int dy \,w^2  (f_r)^2 (f^{(n)}_G)=\frac{1}{3(x^G_n)^4|J_2(x^G_n)|}\int^{x^G_n}_0 dz\, z^3 J_2(z) \\
c_{GG}&=& \int dy \,w^2 (f_r) (f^{(n)}_G)^2=\frac{\sqrt{6}}{4(x^G_n)^4(J_2(x^G_n))^2} \int^{x^G_n}_0 dz\, z^3 (J_2(z))^2.
\eea

For generality, we also include the decay rate of the radion into a pair of KK gravitons, if kinematically allowed,
\bea
\Gamma_r(GG)=\frac{c_{GG}^2\, m_r^3}{32 \pi\Lambda^2 } 
\left(20-\frac{40}{3} \frac{m_r^2}{m_G^2}+\frac{46}{9}\frac{m_r^4}{m_G^4}
-\frac{8}{9}\frac{m_r^6}{m_G^6}+\frac{1}{9}\frac{m_r^8}{m_G^8}\right)
\left(1-\frac{4 m_G^2}{m_r^2}\right)^{\frac{1}{2}}.
\eea

\end{document}